\documentclass[a4paper,fleqn,usenatbib]{mnras}
\usepackage{newtxtext,newtxmath}
\usepackage[T1]{fontenc}
\usepackage{ae,aecompl}
\usepackage{amsmath}
\usepackage{graphicx}	% Including figure files
\usepackage{amssymb}	% Extra maths symbols
\def\betan{\beta_0}

\def\durca/{dUrca}
\def\murca/{mUrca}

\newcommand{\beq}{\begin{equation}}
\newcommand{\eeq}{\end{equation}}
\newcommand{\bea}{\begin{eqnarray}}
\newcommand{\eea}{\end{eqnarray}}

\newcommand{\gv}[1]{\ensuremath{\mbox{\boldmath$ #1 $}}} 
\newcommand{\trm}[1]{\textrm{#1}}

\newcommand{\non}{\nonumber \\}

\newcommand{\grad}[1]{\gv{\nabla} #1}

\newcommand{\be}{\begin{eqnarray}}
\newcommand{\ee}{\end{eqnarray}}
\def\la{\mathbin{\lower 3pt\hbox
      {$\rlap{\raise 5pt\hbox{$\char'074$}}\mathchar"7218$}}}
\def\ga{\mathbin{\lower 3pt\hbox
      {$\rlap{\raise 5pt\hbox{$\char'076$}}\mathchar"7218$}}} %> or of order
\newcommand{\bvec}[1]{\ensuremath{\boldsymbol{#1}}} %boldface vector style
\def\grad{ \mbox{\boldmath $\nabla$} }

\def\kB{k_{\rm B}}

\def\nue{{\nu_{\rm e}}}
\def\nuebar{{\bar{\nu}_{\rm e}}}

\title[Urca reactions during NS inspiral]{Urca reactions during neutron star inspiral}

\author[P. Arras \& N.N. Weinberg]{
Phil Arras,$^{1}$\thanks{E-mail: arras@virginia.edu (PA)}
Nevin N. Weinberg,$^{2}$
\\
$^{1}$Department of Astronomy, University of Virginia, P.O. Box 400325, Charlottesville, VA 22904\\
$^{2}$Department of Physics, and Kavli Institute for Astrophysics and Space Research, \\
Massachusetts Institute of Technology, Cambridge, MA 02139, USA\\
}

\date{Accepted XXX. Received YYY; in original form ZZZ}

\pubyear{2018}

\begin{document}
%\title{$\beta$ disequilibrium during neutron star inspirals}
\label{firstpage}
\pagerange{\pageref{firstpage}--\pageref{lastpage}}
\maketitle

%\author{P. Arras}
%\affil{Department of Astronomy, University of Virginia, P.O. Box 400325, Charlottesville, VA 22904}
%\email{arras@virginia.edu}
%\author{N. Weinberg}
%\affil{Department of Physics, and Kavli Institute for Astrophysics and Space Research, Massachusetts Institute of Technology, Cambridge, MA 02139, USA}
%\email{nevin@mit.edu}

\begin{abstract}

We study the impact of nonlinear bulk viscosity due to Urca reactions driven by tidally-induced fluid motion during binary neutron star inspiral. Fluid compression is computed for low radial order oscillation modes through an adiabatic, time-dependent solution of the mode amplitudes. Optically thin neutrino emission and heating rates are then computed from this adiabatic fluid motion. Calculations use direct and modified Urca reactions operating in a $M=1.4\, M_\odot$ neutron star,  which is constructed using the Skyrme Rs equation of state. We find that the energy pumped into low order oscillation modes is not efficiently thermalized even by direct Urca reactions, with core temperatures reaching only $T \simeq 10^8\trm{ K}$ during the inspiral. Although this is an order of magnitude larger than the heating due to shear viscosity considered by previous studies, it reinforces the result that the stars are quite cold at merger. Upon excitation of the lowest order g-mode, the chemical potential imbalance reaches $\beta \ga 1\, \rm MeV$ at orbital frequencies $\nu_{\rm orb} \ga 200\, \rm Hz$, implying significant charged-current optical depths and Fermi blocking. To asses the importance of neutrino degeneracy effects, the neutrino transfer equation is solved in the static approximation for the three-dimensional density distribution, and the reaction rates are then computed including Fermi-blocking. We find that the heating rate is suppressed by factors of a few for $\nu_{\rm orb} \ga 200\, \rm Hz$. The spectrum of emitted $\nu_e$ and $\bar{\nu}_e$, including radiation transfer effects, is presented for a range of orbital separations.

\end{abstract}


\section{ introduction}

The inspiral of binary neutron stars primarily converts the orbital energy and angular momentum into gravitational waves, as recently observed by the LIGO and VIRGO detectors (GW170817; \citealt{2017PhRvL.119p1101A}). The merger of two neutron stars was also expected to produce an electromagnetic counterpart through the ejection of a small amount of neutron-rich matter which undergoes radioactive decay \citep{1998ApJ...507L..59L,  2010MNRAS.406.2650M}, or the formation of an accretion disk and relativistic jet which produces gamma ray emission \citep{1989Natur.340..126E, 1992ApJ...395L..83N, 2014ARA&A..52...43B}. These counterparts were detected in GW170817 \citep{2017ApJ...848L..12A}.

Tides transfer mechanical energy and angular momentum to the star at the expense of the orbit, and friction within the star then converts the mechanical energy into heat. During the inspiral, these effects are potentially detectable as a deviation of the orbital decay rate from the General Relativistic point-mass result, or as an electromagnetic precursor if heating ejects the outer layers of the star. Different treatments have been used to estimate the transfer of energy and the size of the tidal friction, leading to different conclusions about the importance of pre-merger tidal effects. \citet{1992ApJ...397..570M} parametrized the tidal heating with a fiducial tidal quality factor $Q \sim 10$, finding heating of the core and crust to temperatures $\kB T \gg 1\, \rm MeV$, and mass ejection in a radiation-driven outflow prior to merger. They agreed with the results of \citet{1992ApJ...400..175B} and \citet{1992ApJ...398..234K}, who showed that tidal friction would be unable to spin stars up to corotation before merger for realistic viscosity. Numerous studies \citep{1994ApJ...426..688R,1994MNRAS.270..611L,1999MNRAS.308..153H,2006PhRvD..74b4007L,2017PhRvD..96h3005X} discussed enhanced tidal effects through resonant excitation of g-mode or inertial mode resonances. In particular, \citet{1994MNRAS.270..611L} found that shear and bulk viscosity could only heat the stellar core to energies $\kB T \ll 1\, \rm MeV$. \cite{2017MNRAS.464.2622Y} discuss superfluid effects. While most studies treat the neutron star as a fluid body, \citet{1992ApJ...398..234K} and \citet{2012PhRvL.108a1102T} discuss tides in the solid crust.

%We improve on previous studies of decompression (e.g. \citealt{1977ApJ...213..225L}) through a discussion of neutrino Fermi-blocking effects.\citet{2012PhRvL.108a1102T} discusses possible precursors from ``crust shattering." \citet{1979ApJ...232L.101L}

If tidal friction were efficient during inspiral, and a significant fraction of the tidal energy input to the neutron star was thermalized, intense thermal neutrino emission comparable to that of a supernova would result even before merger. Even if the temperature remains small ($\ll 1\, \rm MeV$), strong fluid compression may result in comparably large neutrino emission before merger, based on the non-equilibrium emissivity given in \citet{1992A&A...262..131H}.  
This is because tide-induced density perturbations drive the fluid out of $\beta$-equilibrium and can induce heating and neutrino emission through Urca reactions. \citet{1994MNRAS.270..611L} discussed ``bulk viscosity" heating by Urca reactions and concluded that the heating is small when $\beta\ll \kB T$ based on the formula from \citeauthor{1989PhRvD..39.3804S} (1989; $\beta$ is the chemical potential imbalance). However, inspiraling NSs are expected to be cold and we find that they are instead in the limit $\beta \gg \kB T$ (a footnote in  \citet{1994MNRAS.270..611L} mentions the $\beta \gg \kB T$ formula in \citealt{1992A&A...262..131H}).  The goal of the present work is to assess the impact of chemical heating and neutrino emission driven by Urca reactions, which have a nonlinear dependence on the tidal perturbations in the limit $\beta \gg \kB T$. In principle, recent numerical simulations of binary neutron inspiral (e.g. \citealt{2003MNRAS.342..673R}) include similar hydrodynamics and neutrino physics as this paper. However, depending on the numerical scheme and resolution, numerical effects which convert kinetic energy to heat may swamp the effects studied here; our results may be used as a code check for high resolution simulations. Further, it is costly to start the simulations at wide separation, and we show that the dominant heating effect during inspiral is due to resonant wave excitation well before merger. 

This work is motivated by Finzi and Wolf (1968, hereafer FW; see also  \citealt{2000A&A...357.1157H} and \citealt{2005MNRAS.361.1415G}), who studied the damping of free oscillations through modified Urca (mUrca) reactions. In a static star in $\beta$-equilibrium, each fluid element's proton fraction will be equal to the equilibrium value, and thermal neutrino emission occurs due to particles within $\kB T$ of the Fermi surface. FW studied the change in neutrino emission due to a compressive fluid motion. The resulting density perturbations cause the proton fraction, approximated as frozen, to deviate from the equilibrium value. The chemical potential imbalance, $\beta$, changes the rates of the \murca/ reactions away from the thermal value, attempting to bring the gas back into $\beta$-equilibrium. The resultant enhancement in neutrino emission, assumed optically thin by FW, acts to cool the star faster than the thermal rate. At the same time, the damping of the mechanical vibration, computed as a thermodynamic $\int P dV$ work integral by FW, acts to heat the star. In the limit $\beta \ll \kB T$, net cooling results, although at a rate slower than the thermal value. However, for $\beta \gg \kB T$, the reaction rates are not only greatly enhanced over the thermal value, but net heating can also occur. The goal of this paper is to apply FW's idea to tidally forced (rather than free) fluid motions during neutron star inspiral. FW's treatment is extended by including tidally forced fluid motions, dUrca as well as mUrca reactions, the derivation of a self-consistent heating rate, as well as radiation transfer and finite optical depth effects such as neutrino Fermi-blocking.
 
Non-equilibrium Urca reactions due to a chemical potential imbalance have found a number of other applications. \citet{1993A&A...271..187G} studied neutrino emission from a neutron star collapsing to a black hole. \citet{1995ApJ...442..749R} studied chemical potential imbalance arising from changes in density profile due to spindown. For that same problem, superfluid effects on the Urca reactions were discussed in \citet{1997ApJ...485..313R}. \citet{2003pasb.conf..231R} discussed ``nonlinear bulk viscosity" damping of free r-mode oscillations in the $\beta \gg \kB T$ limit.

The paper is organized as follows. The heating rate under optically thin conditions is presented in Section \ref{sec:thin}. Section \ref{sec:background} reviews the non-equilibrium Urca reactions and heating rate, as well as the dependence on the equation of state. Section \ref{sec:modeamp} presents the numerical method and estimates for the tidally-induced compression during the inspiral. Section \ref{sec:thinresults} presents results for the chemical potential imbalance, heating rates and core temperature in the optically thin case. Section \ref{sec:transfer} discusses neutrino opacities and the solution of the transfer equation including tidally induced compressions. The opacities in the $\beta \gg \kB T$ limit, and the solution of the transfer equation in the static limit are presented in Section \ref{sec:transferequation}. Reaction rates and emissivities are given in Sections \ref{sec:reactionrates} and \ref{sec:emissivity}. The entropy equation for the n-p-e-$\mu$ gas is discussed in Section \ref{sec:entropy}. Opacities for modified Urca reactions are reviewed in Section \ref{sec:murca}, and compared to the method used in FW. Results for the heating rate and specific neutrino luminosity including neutrino transfer are given in Section \ref{sec:thickresults}. Discussion and conclusions are given in Section \ref{sec:discussion}.

 \section{  the optically thin case  }
 \label{sec:thin}

\subsection{ Background }
\label{sec:background}

\begin{figure}
%\epsscale{2.5}
%\plotone{timescales_Rs_M=1p4.pdf}
\includegraphics[width=\columnwidth]{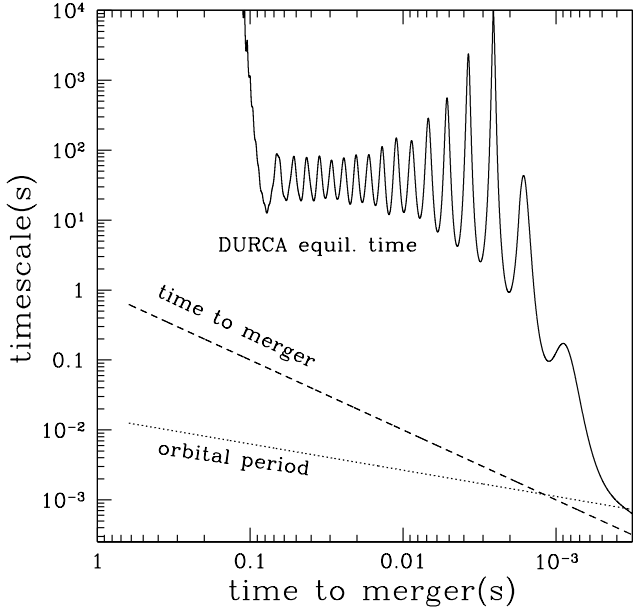}
\caption{\durca/ equilibrium timescale (solid line) versus time to merger (dashed line) and orbital period (dotted line) for the Rs $M=1.4\, M_\odot$ model with a $M^\prime=1.4\, M_\odot$ companion. 
The equilibration timescale  is always much longer than either the time to merger or the orbital period. The oscillations  at $10^{-1}-10^{-2}\rm s$ are due to the resonant excitation of the g$_1$ mode. Between $10^{-2}-10^{-3}\rm s$ the rapid decrease in the equilibration time is due to the non-resonant response of the f-mode.}
\label{fig:equil_time}
\end{figure}

When the star is in $\beta$-equilibrium, the chemical potential imbalance 
\beq
\beta\equiv \mu_n -\mu_p-\mu_e = 0,
\eeq
where $\mu_n$, $\mu_p$, and $\mu_e$ are the chemical potentials of the neutron, proton, and electron, respectively (we assume the neutron star is composed entirely of neutrons, protons, electrons, and muons).
Deviations from $\beta$-equilibrium induce electron-type neutrino and antineutrino \footnote{ Urca reactions involving $\mu$ neutrinos and antrineutrinos are ignored for simplicity. They would contribute at the factor of 2 level at high density where the muons are relativistic.} emission through the direct Urca (\durca/) process 
\beq
n\rightarrow p + e + \bar{\nu}_e, \hspace{0.6cm} p+e\rightarrow n+\nu_e,
\label{eq:dUrca}
\eeq
and the modified Urca (\murca/) processes
\bea
n+n\rightarrow p+n+e+\bar{\nu}_e,&& \hspace{0.4cm} p+n+e\rightarrow n+n+\nu_e;\hspace{0.5cm}\nonumber \\
n+p\rightarrow p+p+e+\bar{\nu}_e,&& \hspace{0.4cm} p+p+e\rightarrow n+p+\nu_e
\label{eq:mUrca}
\eea
(see \citealt{2001PhR...354....1Y}).  At low temperatures $\kB T \ll |\beta |$, the reactions on the left hand side run when $\beta>0$ and the reactions on the right hand side run when $\beta < 0$.  The \durca/ process only occurs if the Fermi momenta satisfy $p_{Fn} < p_{Fp}+p_{Fe}$, which can be shown to correspond to a critical proton fraction $x_p \ga 1/9$ \citep{1991PhRvL..66.2701L}.

Density perturbations induced by the tide drive fluid elements out of $\beta$-equilibrium such that
\beq
\beta(\gv{x},t) \simeq \betan(r) \left(\frac{\Delta\rho(\gv{x},t)}{\rho(r)}\right),
\label{eq:beta}
\eeq
where $\rho$ is the background density, $\Delta \rho$ is the Lagrangian (comoving) density perturbation, and the thermodynamic derivative $\betan \equiv (\partial \beta/\partial \ln \rho)_{s,x}$. Here we assume that the perturbations are small (linear) and that only the fluid element's density varies, not its entropy $s$ or the concentrations $x=(x_{\rm p}, x_{\rm e}, x_{\rm n},x_{\rm \mu})$. The latter assumption is reasonable because the equilibration timescale is much longer than the orbital period and time to merger during the inspiral (see Fig. \ref{fig:equil_time} and Section \ref{sec:reactionrates}). In this limit, we can estimate the heating rate by computing the adiabatic response of the star to the tidal force and then plugging this response into the reaction rates.

\begin{figure}
%\epsscale{2.5}
%\plotone{Rs_M=1p4_model.pdf}
\includegraphics[width=\columnwidth]{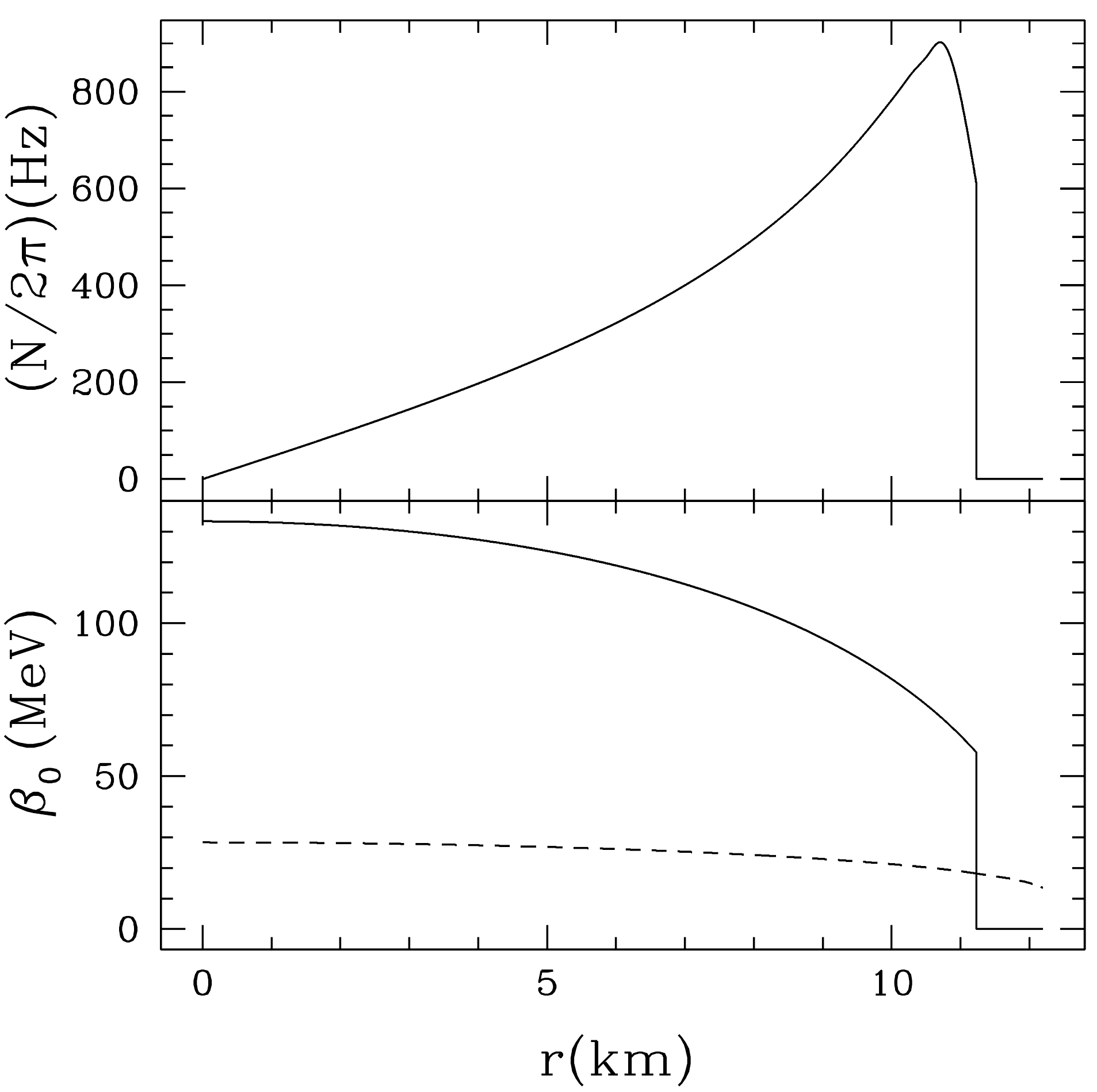}
\caption{Profile of the Brunt-Vaisalla frequency (top)  and $\betan$ (bottom) versus radius. The solid line is the Rs model model including nuclear interactions and the dashed line is the noninteracting result $\betan \simeq \mu_e/3$. }
\label{fig:model}
\end{figure}

In the limit of noninteracting nucleons, $\betan \simeq \mu_e/3\simeq20\, {\rm MeV} (\rho/\rho_{\rm nuc})^{1/3}$ \citep{1992ApJ...395..240R}, where nuclear density $\rho_{\rm nuc}\simeq 2.7\times10^{14}\trm{ g cm}^{-3}$.
However, considerably larger values are possible for interacting nuclear matter. The lower panel of Fig. \ref{fig:model} shows that for a $M=1.4\, M_\odot$ model including interactions, $\betan\sim 100\trm{ MeV}$ over much of the core.  In Section \ref{sec:modeamp} we show that in the late inspiral, $\Delta \rho/\rho \sim 0.01$ (see Fig. \ref{fig:Drho}) and hence $\beta \sim 1 \trm{ MeV}$. 

We use the Skyrme interaction  with Rs model parameters for the equation of state \citep{1997NuPhA.627..710C}. We choose the Skyrme parametrization because it allows for a straightforward computation of all the needed thermodynamic derivatives.  We specifically choose the Rs model  because it has a symmetry energy that increases relatively strongly with density.  As a result,  the proton fraction $x_p>1/9$ at relatively low $\rho$ and the direct Urca process can operate in the core even for $M=1.4 M_\odot$.  However, as Table VI in \citet{2003PhRvC..68c4324R} shows, the Rs model is not unique in this regard among viable Skyrme models.  

There is observational evidence that the \durca/ process might be active in some NSs.  For example, rapid cooling due to the \durca/ process could explain the low luminosity of several young supernova remnants \citep{Kaplan:04, Kaplan:06, Shternin:08}.  It could also explain why the pulsar in CTA1, the transiently accreting millisecond pulsar SAX J1808.4-3658, and the soft X-ray transient 1H 1905+000 appear to be very cold \citep{Jonker:07, Heinke:09, Page:09}.  On the other hand, \citet{Klahn:06} argue that constraints on the NS equation of state do not allow the \durca/ process to occur in NSs with masses below $1.5M_\odot$.

\begin{figure}
\includegraphics[width=\columnwidth]{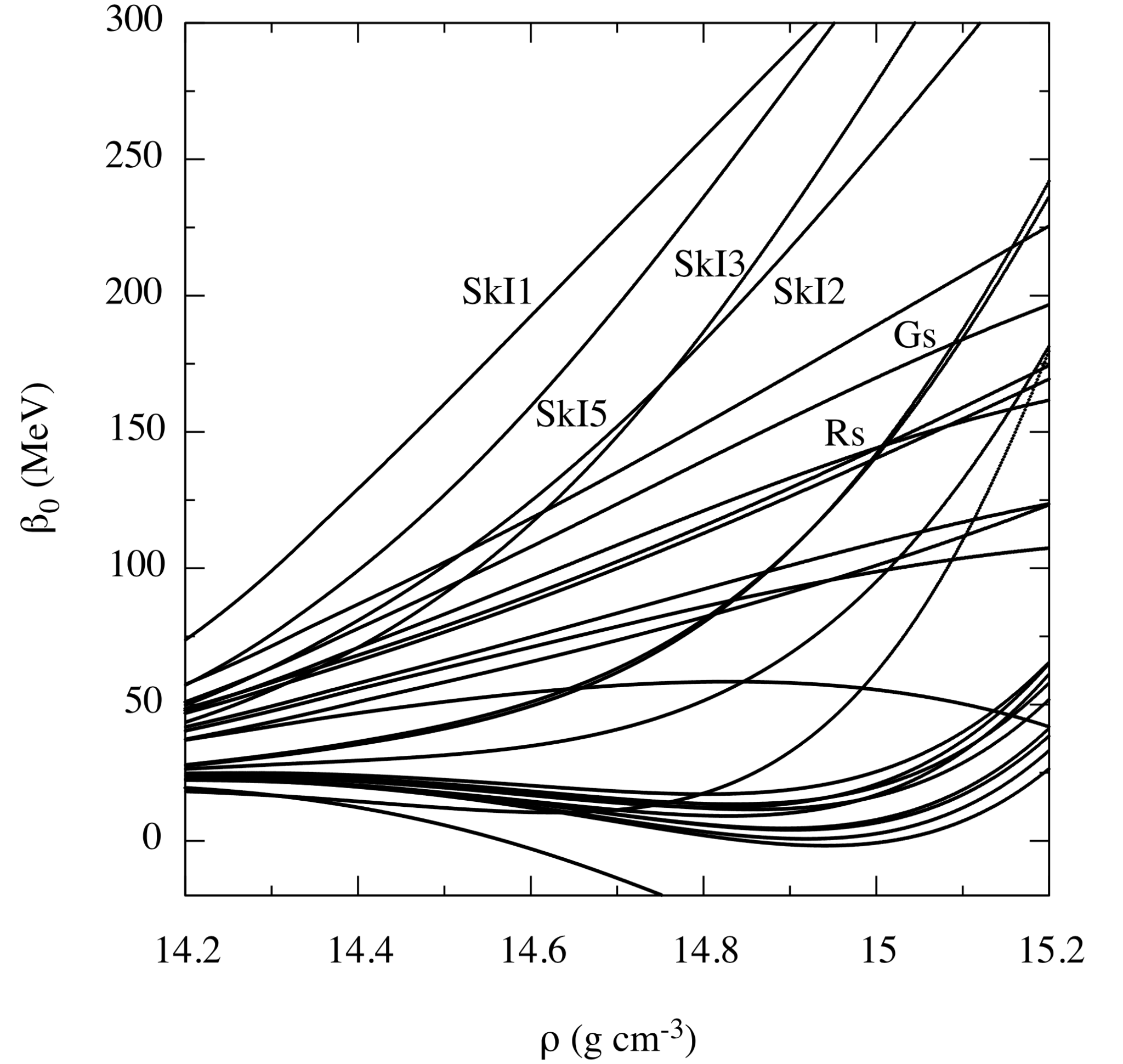}
%\hspace*{-1.8cm}
%\includegraphics[width=9.0cm]{beta0_EOS.png}
\caption{$\beta_0$ versus density $\rho$ for the set of Skyrme parameterizations recommended by \citet{2003PhRvC..68c4324R} as they yield neutron star models that are consistent with observed constraints. The parameterizations with the largest $\beta_0$ are labeled.  We use the Rs model throughout the paper.}
\label{fig:beta0_EOS}
\end{figure}

The value of $\beta_0$ is sensitive to the equation of state. In Figure~\ref{fig:beta0_EOS} we show $\beta_0$ as a function of density $\rho$ for Skyrme parameters that \citet{2003PhRvC..68c4324R} find give satisfactory neutron star models (their Table VII).  While the Rs model has a relatively large $\beta_0(\rho)$, the SkI models have values of $\beta_0$ that are more than twice as big.  As we show below (see Equation~\ref{eq:Edotheat}), the heating rate $\dot{E}_{\rm heat}$ depends on $(\beta_0 \Delta \rho /\rho)^6$. A model with even a slightly larger $\beta_0$ might have a significantly larger $\dot{E}_{\rm heat}$, although in practice we find that this effect may be mitigated by a decrease in $\Delta \rho /\rho$ with increasing $\beta_0$.

Inspiraling neutron stars are expected to be colder than the critical temperature for proton superconductivity and neutron superfluidity $T_c \simeq 10^8-10^{10}\trm{ K}$ \citep{2003RvMP...75..607D}.  For $\beta \ll \kB T$, superfluidity and superconductivity exponentially suppress the rate of Urca processes \citep{2001PhR...354....1Y}.  However, for $\beta \ga \kB T_c \gg \kB T$, there is very little suppression \citep{2010PhRvC..81d5802P, 2012PhRvL.108k1102A}.   Since we find that $\beta \sim 1 \trm{ MeV} \ga \kB T_c$ for $\nu_{\rm orb}\ga 100\trm{ Hz}$, the Urca reaction rates and emissivities should approximately equal those of a normal fluid during the late inspiral.  Furthermore, \citet{2017MNRAS.464.2622Y,2017MNRAS.470..350Y} showed that superfluidity has only a mild influence on a neutron star's tidal response (dynamical and equilibrium) during inspiral.  For simplicity, we therefore treat the neutron star as a normal fluid.

Assuming that the core temperature $\kB T \ll \beta \sim 1 \trm{ MeV}$, the net heating rate  (chemical heating rate minus neutrino cooling rate)  due to the \durca/ process can be shown to equal the neutrino emission rate (see Section \ref{sec:entropy}), and is given by (\citealt{1992A&A...262..131H} and Section \ref{sec:emissivity})
\be
\dot{E}_{\rm heat} & \simeq &
\int d^3x\, \varepsilon\, \left( \frac{\Delta \rho}{\rho} \right)^6
\label{eq:Edotheat}
\ee
where the emissivity
\be
\varepsilon & \simeq &  4.5 \times 10^{29}\, {\rm erg\ cm^{-3}\ s^{-1}}\, 
\Xi \left( \frac{n_e}{n_0} \right)^{1/3} \left( \frac{\betan}{1\, \rm MeV} \right)^6.
\label{eq:varepsilon_H92} 
\ee
and the linear approximation for $\beta$ in equation~(\ref{eq:beta}) was used. Here $n_e$ is the electron density and $n_0=0.16\, \rm fm^{-3}$ is the nuclear saturation density. The factor $\Xi=1$ if $p_{Fn} < p_{Fp}+p_{Fe}$ and $\Xi=0$ otherwise. Since the heating rate scales as  $\Delta \rho^6$, it increases dramatically during the inspiral as the tidal force increases.

\subsection{Mode amplitudes and $\Delta \rho$}
\label{sec:modeamp}

The ``equilibrium tide" denotes an approximation to the fluid motion in which inertia, the $d\gv{v}/dt$ term in the fluid equations, is ignored. In this limit the fluid responds instantaneously to the tidal forcing. This approximation is often used as it admits a simple, analytic solution to the fluid equations (e.g. \citealt{1989ApJ...342.1079G}). However, the linear equilibrium tide has zero fluid compression, and hence cannot drive the Urca reactions. 

The ``finite-frequency equilibrium tide" \citep{2010ApJ...714....1A,2016ApJ...819..109W} includes compression by treating the tidal forcing frequency, $\omega=4\pi\nu_{\rm orb}$, as a small parameter and applying perturbation theory. With this approach, the compression scales as 
\be
\frac{\Delta \rho_{\rm ff}}{\rho} & \sim & \left( \frac{\omega^2}{N^2} \right) \left( \frac{U}{c_s^2} \right),
\label{eq:Drhoff}
\ee
where $N^2$ is the Brunt-Vaisalla frequency, $c_s$ is the sound speed and $U$ is the tidal potential.
Since $U \propto \omega^2$, equation~(\ref{eq:Drhoff}) gives a dUrca heating rate $\dot{E}_{\rm heat} \propto \nu_{\rm orb}^{24}$, which rises very steeply toward smaller separation. For stable stratification due to composition gradients, $N^2 \propto \beta_0$ and hence $\beta_0 \Delta \rho_{\rm ff}/\rho$ is independent of $\beta_0$ for the finite frequency equilibrium tide. While equation~(\ref{eq:Drhoff}) is a simple analytic formula, it is not numerically accurate in the present case due to the resonant excitation of oscillation modes.

The ``dynamical tide" represents a wavelike response to the tidal force. We account for the contributions of the quadrupolar f-mode (f), the lowest order p-mode (p$_1$) and the two lowest order g-modes (g$_1$ and g$_2$) to $\Delta \rho$. The f-mode couples strongly to the tidal potential but is non-resonant and weakly compressive. It has zero compression in a constant density, incompressible stellar model but acquires a small compression in realistic stellar models. The g-modes couple relatively weakly to the tidal potential but undergo resonant excitation as the orbital frequency sweeps upward during the inspiral. Even for incompressive motions, the exact dynamical tide solution is only well represented by the equilibrium tide approximation when the forcing frequency is lower than the frequency of low radial order modes, in this case the g$_1$ and $f$-modes.

The total Lagrangian displacement vector of the adiabatic fluid response can be written as a sum over normal modes 
\be
\bvec{\xi}\left( \bvec{x},t \right) & = & \sum_\alpha q_\alpha(t) \left[ \xi_{r,\alpha}(r) Y_{\ell m}(\theta,\phi) \bvec{e}_r
\right. \nonumber \\ & + & \left. \xi_{h,\alpha}(r) r \grad Y_{\ell m}(\theta,\phi) \right],
\label{eq:mode_expansion}
\ee
where $q_\alpha(t)$ is the time-dependent, dimensionless mode amplitude, $\xi_{r,\alpha}(r)$ and $\xi_{h,\alpha}(r)$ are the radial and horizontal displacements, and $Y_{\ell m}(\theta,\phi)$ is the spherical harmonic of order $\ell$ and index $m$.  Each mode is labeled by $\alpha=\{\ell, m, \omega_\alpha\}$, where  the eigenfrequency $\omega_\alpha$ depends on the number of radial nodes $n_\alpha$ and the mode type (p-mode, g-mode, or f-mode). In addition, for each mode with $(m,\omega_\alpha)$ there is a ``complex conjugate mode" with the same $\ell$ and $(-m,-\omega_\alpha)$. The mode and its complex conjugate have the same phase velocity $\omega_\alpha/m$ and represent the ability to express the $\phi$ and $t$ dependence in terms of sines and cosines. 

Fig. \ref{fig:model} (upper panel) shows the Brunt-Vaisalla frequency
\be
N^2 &= & - g \left( \frac{d\ln\rho}{dr} - \frac{1}{\Gamma_1} \frac{d\ln P}{dr} \right),
\ee
which sets the frequencies of g-modes, for the Rs $M=1.4\, M_\odot$ model. For this zero-temperature model, stable stratification is due to composition gradients \citep{1992ApJ...395..240R}.
Here $P$ is the pressure, $\Gamma_1 = (\partial \ln P/\partial \ln \rho)_{s,x}$ is the first adiabatic index, and $g=Gm/r^2$ is the gravity. The Brunt-Vaisalla frequency has been set to zero in the crust, for simplicity, following \citet{1994MNRAS.270..611L}. 

We normalize the modes such that $|q_\alpha|=1$ corresponds to a mode energy $E_0=GM^2/R=4.2\times10^{53}\trm{ erg}$ for our neutron star model
with mass $M=1.4\, M_\odot$, radius $R=12\, \rm km$ and dynamical frequency $\nu_0=\sqrt{GM/R^3}/(2\pi)=1600\, \rm Hz$. Properties of the low order eigenmodes of the Rs $M=1.4\, M_\odot$ model are given in Table \ref{table:emodes}. Although we construct the background neutron star by solving the TOV equations, for simplicity we solve for the eigenmodes using the Newtonian oscillation equations. This may introduce $\sim50\%$ errors in the eigenfrequencies and tidal overlap integrals \citep{1994ApJ...432..296R}.

The heating rate is expanded as a sextuple sum over spherical harmonics
\be
\dot{E}_{\rm heat}(t) &\simeq & 
\sum_{\ell_1m_1} \cdots \sum_{\ell_6m_6} \left( \int d\Omega Y_{\ell_1 m_1} \cdots Y_{\ell_6 m_6} \right)
\nonumber \\ & \times &
\left(  \int_0^R dr r^2 \epsilon(r)
 \frac{\Delta \rho_{\ell_1 m_1}(r,t)}{\rho(r)} \cdots \frac{\Delta \rho_{\ell_6m_6}(r,t)}{\rho(r)} \right)
\label{eq:Edotheatlm}
\ee
where the spherical harmonic coefficient is given by a sum over modes with that $\ell$ and $m$ as
\be
\Delta \rho_{\ell m}(r,t) & = & \sum_\alpha q_\alpha(t) \Delta \rho_{\alpha}(r)
\label{eq:Delta_rho}
\ee
and the eigenfunction is computed as
\be
 \frac{\Delta \rho_{\alpha}}{\rho} & = & \frac{\delta p_{\alpha}/\rho - g \xi_{r,\alpha}}{c_s^2},
\ee
where $\delta p_{\alpha}(r)$ is the Eulerian pressure perturbation, $\xi_{r,\alpha}$ is the radial component of the displacement vector, and 
$c_s^2=\Gamma_1 P/\rho$ is the adiabatic sound speed. The linear oscillation equations for $\delta p_\alpha$ and $\xi_{r,\alpha}$, including the perturbation to the gravitational potential perturbation, are solved as in \citet{2012ApJ...751..136W}.

The requirement that $\Delta \rho(\bvec{x},t)$ be real implies 
$\Delta \rho^*_{\ell m}=(-1)^m\Delta \rho_{\ell-m}$.  
The angular integral is only nonzero for $m_1+\cdots+m_6=0$. 
For example, integrals involving only $|m|=2$ must contain three factors of $m=2$ and three factors of $m=-2$. There are 20 such permutations, each one giving an integral $\int d\Omega |Y_{22}|^6 = 1125/(8008\pi^2)$. 
Integrals containing 2, 4 or 6 of the $Y_{20}$ harmonics are also possible. 
%\nevin{ should we say anything about $\ell>2$ here? or later?} \phil{$\ell=3$ run up and running. add results or just state?}

\begin{table}
\centering
\caption{Properties of the low-order eigenmodes of the $M=1.4\,M_\odot$ neutron star model}
\begin{tabular}{ccccc}
%\tablehead{
%\colhead{mode id} & \colhead{$\ell$} & \colhead{$n_\alpha$}  & \colhead{$\omega_\alpha/2\pi$(Hz)} & \colhead{$I_{\alpha\ell m}$}
%}
%\startdata
\hline
mode id & $\ell$ & $n_\alpha$ & $\omega_\alpha/2\pi$(Hz) & $I_{\alpha \ell m}$ \\
\hline
$p_1$ 	&	2 	&	1	&   $8.27\times 10^3$	&	$1.81 \times 10^{-3}$	\\
$f$      	& 	2	&	0	&   $2.09\times 10^3$	&	$2.81 \times 10^{-1}$	\\
$g_1$   & 2	&	1	&   $3.16\times 10^2$	&	$9.23\times 10^{-3}$		\\
 $g_2$  & 2	&	2	&   $1.93\times 10^2$	&	$1.83\times 10^{-3}$	\\
 \hline
\end{tabular}
%    \enddata
    %\tablecaption{text\label{key}}
    \label{table:emodes}
\end{table}

To determine the $q_{\alpha}(t)$,    let the primary, tidally forced star have mass $M$ and radius $R$, and the perturber have mass $M^\prime$. The primary is assumed to be non-rotating as tidal dissipation has been estimated to be too weak to maintain synchronous rotation during inspiral \citep{1992ApJ...400..175B,1992ApJ...398..234K}.
In a spherical coordinate system $(r, \theta, \phi)$ centered on the primary, we take the orbit of the perturber to be $(D(t), \pi/2, \Phi(t))$, where D(t) is the separation and $\Phi(t)$ is the orbital phase. Since tidal interactions and gravitational wave emission by the induced stellar quadrupoles have a small overall effect on the inspiral, we use the quadrupole formula for the rate of orbital decay of two point masses \citep{1964PhRv..136.1224P}:
\beq
\dot{D}=-\frac{64G^3}{5c^5}\frac{MM'(M+M')}{D^3},
\hspace{0.3cm} \dot{\Phi}=\left[\frac{G(M+M')}{D^3}\right]^{1/2}.
\label{eq:dotDgw}
\eeq
The tidal potential due to the companion is
\be
U(\bvec{x},t)  & = &  -  GM^\prime
\sum_{\ell \geq 2,m} 
\frac{W_{\ell m}  r^\ell}{D^{\ell+1}(t)} Y_{\ell m}(\theta,\phi) e^{-i m\Phi(t)},\hspace{0.2cm}
\ee
where $W_{\ell m} =4\pi (2\ell+1)^{-1} Y_{\ell m}(\pi/2,0)$. The evolution equation for the mode amplitude $q_{\alpha}(t)$ due to driving by the linear tidal force is then  \citep{2002PhRvD..65b4001S,1994MNRAS.270..611L, 2012ApJ...751..136W}
\be
\dot{q}_\alpha + i \omega_\alpha q_\alpha &= & i \omega_\alpha U_\alpha(t),
\label{eq:amp_eqn}
\ee
where
\bea
U_\alpha(t) & = & -\frac{1}{E_0} \int d^3x \rho \bvec{\xi}^*_\alpha(\bvec{x}) \cdot \grad U(\bvec{x},t).
\non & = & \frac{M^\prime}{M}  
W_{\ell m}I_{\alpha \ell m}  \left( \frac{R}{D(t)} \right)^{\ell+1}  e^{-i m \Phi(t)},
\eea
and the dimensionless overlap integral 
\be
I_{\alpha \ell m} & =& \frac{1}{M R^\ell} \int d^3x \rho \,\bvec{\xi}^*_\alpha(\bvec{x}) \cdot \grad\left(r^\ell Y_{\ell m} \right).
\ee
The overlap integral is a normalization-dependent quantity. In the limit that the tidal forcing frequency is much smaller than the mode's natural frequency, $\omega_\alpha$, the simple ``equilibrium tide" solution for the mode amplitude is $q_\alpha(t) \simeq U_\alpha(t) \propto I_{\alpha \ell m}$.

\begin{figure}
%\epsscale{2.5}
%\plotone{Lag_drho_vs_r.pdf}
\includegraphics[width=\columnwidth]{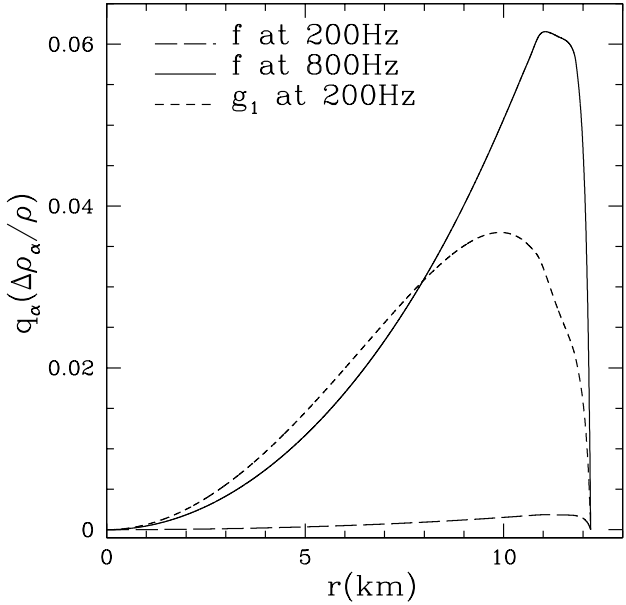}
\caption{Comparison of the Lagrangian density perturbation for the $\ell=2$ f-mode and g$_1$-mode. The f-mode eigenfunction has been multiplied by $-1$ so that it may be more easily compared to the g$_1$ eigenfunction.}
\label{fig:Drho}
\end{figure}

Fig. \ref{fig:Drho} shows the run of $q_\alpha \Delta \rho_{\alpha \ell}(r)/\rho(r)$ over the star for the f-mode and g$_1$ mode. Aside from the $Y_{\ell m}(\theta,\phi)$, $q_\alpha \Delta \rho_{\alpha \ell}(r)/\rho(r)$ is the physical density perturbation. The f-mode result is displayed for two different times during the inspiral, with the mode amplitudes found in Fig. \ref{fig:mode_energies}. The g-mode result is displayed at the resonant amplitude found above $\nu_{\rm orb} \simeq 200\, \rm Hz$.  In spite of the much larger overlap integral for the f-mode, it is so weakly compressive that the g$_1$ mode is comparable in the core ($r \la 7\, \rm km$), where dUrca is active, until late in the inspiral (see Fig. \ref{fig:Edot}). The compression has a characteristic size $\Delta \rho/\rho \sim 10^{-2}-10^{-1}$ near the merger.

We calculate $q_{\alpha}(t)$ by solving equation~(\ref{eq:amp_eqn}) for each mode as a function of time starting from large separation up to merger.   The f-mode's frequency is too large for it to ever be resonant with the tidal forcing, since for this model $\omega_{f22}/2\pi =2090\trm{ Hz}$ whereas $D=2R$ at $2\dot{\Phi}=1600\trm{ Hz}$.  Therefore, the f-mode always oscillates at the tidal forcing frequency $2\dot{\Phi}$ and its energy increases with time as
\bea
E_{\trm{f}}(t)&\simeq&  \left( \frac{\omega_f}{\omega_f-2\dot{\Phi}} \right)^2 \left[\frac{M'}{M} \left(\frac{R}{D(t)}\right)^3 I_{\trm{f}22}\right]^2 E_0
\non &\approx & 5\times10^{49} \left( \frac{\omega_f}{\omega_f-2\dot{\Phi}} \right)^2 \left(\frac{M'}{M}\right)^2 \left(\frac{3R}{D(t)}\right)^{6}\trm{ erg},
\eea
where the Lorentzian factor may reach $\ga 10$ by merger. The p-modes have higher frequency than the f-mode but much smaller overlaps, and give a negligible contribution to the energy.
The g-modes are resonantly excited as the binary sweeps through their eigenfrequencies \citep{1994ApJ...426..688R, 1994MNRAS.270..611L,1994PThPh..91..871S}.  Before resonance, their amplitudes are small and they oscillate at frequency $2\dot{\Phi}$ .  After resonance, they oscillate at their natural frequency.  They are excited to a maximum energy that is set by the competition between their driving rate at resonance and the duration of the resonance. Solving equation~(\ref{eq:amp_eqn}) using the stationary phase approximation gives \citep{1994MNRAS.270..611L} 
\bea
E_{\trm{g},\trm{max}}&\simeq& 
\frac{\pi^2 \kappa}{256}\left(\frac{GM}{Rc^2}\right)^{-5/2}\left(\frac{\omega_\alpha}{\omega_0}\right)^{7/3}I_{\alpha 22}^2 E_0
\non&\approx& 
2.4\times10^{48} \kappa \left(\frac{\omega_{\trm{g}}/2\pi}{316\trm{ Hz}}\right)^{7/3}\left(\frac{I_{\alpha 22}}{0.009}\right)^2\trm{ erg},\hspace{0.3cm}
\eea
where $\kappa=(M'/M)\left[2M/(M+M')\right]^{5/3}$ and $\omega_0=(GM/R^3)^{1/2}$ is the primary's dynamical frequency.  In the second line, we plugged in values corresponding to the $g_1$ mode (the higher order modes have smaller $E_{\trm{g},\trm{max}}$). The maximum tidal energy input occurs for $M' \sim M$ because there is then a balance between the tidal amplitude and time spent in resonance.

\subsection{ Results }
\label{sec:thinresults}

In this section results are given in the optically thin limit for the $M=1.4\, M_\odot$ tidally forced star with a $M^\prime=M=1.4\ M_\odot$ companion. There are 24 modes included, the four modes given in Table \ref{table:emodes}, augmented with $m=-2,0,2$ and both signs of the frequency. The starting orbital frequency is taken to be $\nu_{\rm orb}=80\, \rm Hz$ ($D\simeq 10 R$), well below any eigenfrequency, and the integration is allowed to extend to (unphysically) large frequencies at thousands of Hz. In the figures, we show two physically motivated conditions for the breakdown of the linear approximation: (i) when the stars come into Roche lobe contact \citep{1983ApJ...268..368E}, and (ii) when the stars merge at $D=2R$, which is near the ISCO (i.e,  $6G(M+M')/c^2 \simeq 2R$).

\begin{figure}
%\epsscale{2.5}
%\plotone{mode_energies_1p4_1p4.pdf}
\includegraphics[width=\columnwidth]{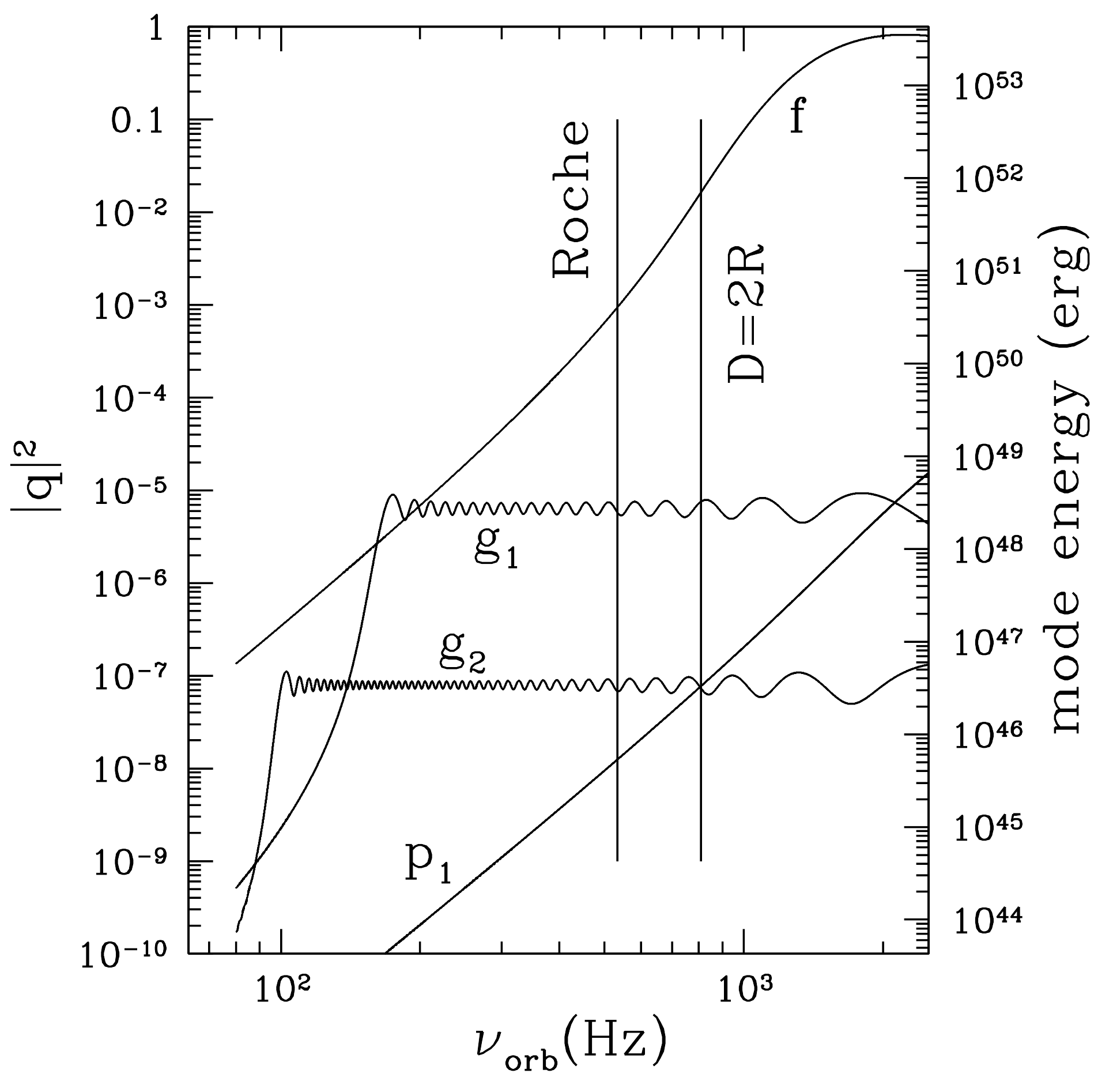}
\caption{Mode energies in dimensionless form (left axis) and in units of erg (right axis) during inspiral. Each mode is labeled as acoustic (p$_1$), fundamental (f), lowest order g-mode (g$_1$) and a higher radial order g-mode (g$_2$). The parameters of each mode are given in Table \ref{table:emodes}. The vertical lines show the point at which the two unperturbed stars touch (right) and when each star overfills its Roche lobe (left). }
\label{fig:mode_energies}
\end{figure}

Fig. \ref{fig:mode_energies} shows the amplitudes of prograde modes which may have a resonance with the tide. The p$_1$ and f modes are not resonant during the inspiral before merger. The two g-modes are pumped by the tide at resonance, with g$_1$ having much larger energy than g$_2$. Higher order p and g-modes are unimportant for heating. The f-mode response, which is not yet resonant even when the two (background, non-tidally distorted) stars are touching, has larger energy than the g-modes except very near resonances.  Retrograde and $m=0$ modes, not shown, do not have resonances, but do have energy pumped in by the tide. They are also included in the heating rate.

\begin{figure}
%\epsscale{2.5}
%\plotone{Edot_vs_nu_1p4_1p4.pdf}
\includegraphics[width=\columnwidth]{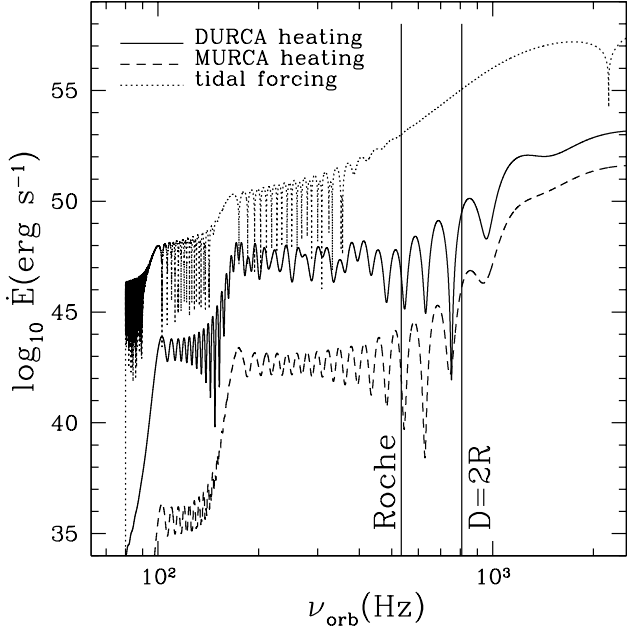}
\caption{Energy input versus orbital frequency. The dotted line shows the rate at which mechanical energy is pumped into the oscillation modes by the tidal force. The solid line shows the rate at which that mechanical energy is damped into heat by the \durca/ process. The dashed lines shows the rate of heating by the \murca/ reaction.  }
\label{fig:Edot}
\end{figure}

Fig. \ref{fig:Edot} shows the rate at which orbital energy is input into oscillation modes by the tide, and the rate at which the oscillation mode energy is damped to heat the star. The damping rates are always small compared to the tidal pumping rate, since the Urca equilibration rates are so slow. While the mode amplitude equations ignored damping, even if damping was included it would have only caused minor decrease in the amplitude of the g$_1$ and g$_2$ modes between the resonance time and merger. The large oscillations seen in the \durca/ (solid) and \murca/ (dashed) lines are due to the pumped g-modes. The portions where relatively steady increases are found are due to the non-resonant f-mode response.  Even though the g$_1$ mode energy is much smaller than the f-mode energy throughout the late inspiral  (see Fig. \ref{fig:mode_energies}), it contributes significantly to the heating due to its larger compressibility (see Fig. \ref{fig:Drho}). 
\begin{figure}
%\epsscale{2.5}
%\plotone{beta_T_vs_nu_1p4_1p4.pdf}
\includegraphics[width=\columnwidth]{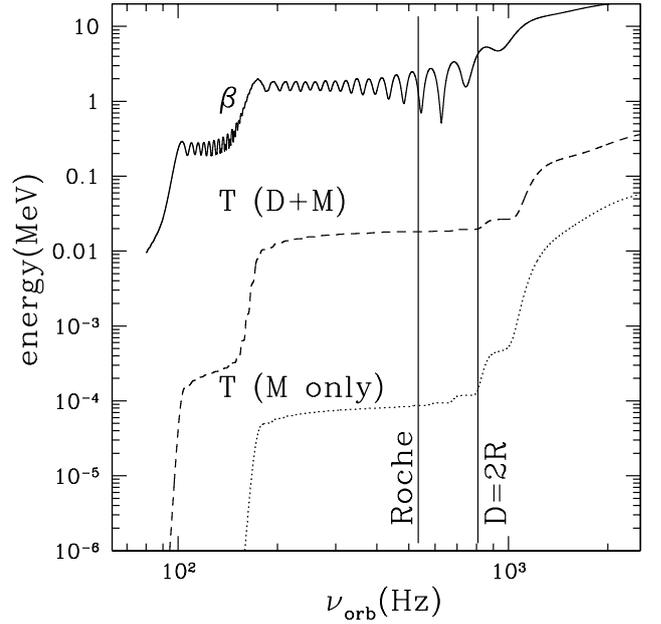}
\caption{Chemical potential imbalance ($\beta$, solid line) and temperature ($T$) in units of MeV. The dashed line is the temperature including both \durca/ and \murca/ heating, and the dotted line only includes \murca/ heating.}
\label{fig:betaT}
\end{figure}

Fig. \ref{fig:betaT} shows the tidally induced chemical potential imbalance ($\beta$) and temperature due to accumulated heat input during the inspiral. The temperature is found by equating the accumulated heat input to the total thermal energy (assumed to be from normal neutrons) of the core, and represents an average over the core \footnote{ The heating may be concentrated in both latitude and radius, so that a core average may underestimate the highest temperatures achieved.}. Two cases are shown, with the sum of \murca/ and \durca/ heating, and only including \murca/ heating. As expected, \durca/ dominates by many orders of magnitude, especially at wide separations since \murca/ scales as $\beta^8$ while \durca/ scales as $\beta^6$. The value of $\beta$ shown is given by the maximum of the angular integral $(\int d\Omega \beta^6/4\pi)^{1/6}$ over the radial grid at each orbital separation. 
In the pre-merger phase, the chemical potential imbalance can reach values of $\beta \sim 1\, \rm MeV$ with values $\beta \sim 5\, \rm MeV$ associated with f-mode excitation at merger. Optical depth $\tau > 1$ is a concern at $\beta \ga 0.3\, \rm MeV$; this will be discussed in detail in Section \ref{sec:transfer}. We find that \durca/ heating can heat the core to temperatures $T \simeq 2\times 10^{-2}\, \rm MeV$ ($2\times 10^8\, \rm K$) before merger.
By comparison, \murca/ heating is always negligible. A core temperature $T \simeq 10^{-3}\ \rm MeV$ was found for shear viscosity damping of the equilibrium and dynamical tides (\citealt{1994MNRAS.270..611L}, with a typo corrected, as discussed in  \citealt{2013ApJ...769..121W} and \citealt{2017MNRAS.464.2622Y}).

Only one equation of state (Skyrme Rs) has been used for the results shown here. Experimentation with other Skyrme parameters (e.g. SLy4) showed that the Rs equation of state has two aspects which promote large Urca rates. First, the rapid increase of the symmetry energy with density leads to large proton fractions at high density, and hence large core g-mode frequencies, where buoyancy is due to the proton gradient. Second, the proton fraction is large enough to allow \durca/ reactions, even for this $M=1.4\, M_\odot$ star. If \durca/ is not allowed, the results show that \murca/ heating is much weaker.

\section{radiative transfer effects} 
\label{sec:transfer}

In the late stages of the inspiral, the chemical potential imbalance ($\beta$) is much larger than the thermal energy ($\kB T$; see Fig. \ref{fig:betaT}). Deviations from $\beta$ equilibrium cause neutrino emission through the \durca/ and \murca/ reactions (Eqs. \ref{eq:dUrca} and \ref{eq:mUrca}). Here the $T=0$ limit of the transfer equation is studied in which the sources and sinks of radiation depend on the dynamically generated chemical potential imbalance. Finite charged-current optical depth effects are included, and their effects on heating are compared to the optically thin case. Formulae in this section will use $\hbar=\kB=c=1$ units, with numerical estimates given in CGS units.

In addition to the Urca reactions, another source of opacity is neutral-current neutrino scattering off neutrons. If scattering optical depths were to become large,  it would imply a significant lag between neutrino emission and escape from the star, and the static approximation that we make below may no longer be valid. However, we find that the scattering mean free paths should not become large during the inspiral.  Using the result from \cite{1979ApJ...230..859S},  the scattering mean free path in the $T=0$ limit is
\be
\lambda_{\rm \nu-n} & \simeq & 10^3\, \rm km\, \left( \frac{1\, \rm MeV}{k} \right)^3 \left( \frac{\rho_{\rm nuc}}{\rho} \right)^{2/3},
\label{eq:nunmfp}
\ee
where $k$ is the neutrino energy. This formula applies for degenerate neutrons and also requires $2kv_{\rm F,n} \gg T$, which is the case here for $k \sim \rm MeV$, $v_{\rm F,n} \sim 0.3$ and $T \sim 10^{-2}\, \rm MeV$.
The mean free path in equation~(\ref{eq:nunmfp}) is longer than that for scattering off non-degenerate neutrons by a factor $p_{\rm F,n}/T \gg 1$. In order for 
the scattering mean free path to be smaller than the size of the star requires high energy neutrinos $k \ga 5\, \rm MeV$ at nuclear density, which occurs only at orbital separations approaching merger. Hence our neglect of neutral current scattering to simplify the problem is approximately valid throughout the inspiral. 

Neutral current elastic scattering of neutrinos with heavy nuclei in the crust has been estimated by \citet{1993A&A...271..187G} to have an optical depth $\tau \sim 1\, (k/\rm MeV)^2(A/100)$ for nucleon mass $A$. This will also aid to trap neutrinos with energies $k \ga 1\, \rm MeV$ near merger. This effect is ignored for simplicity.

\subsection{ The transfer equation }
\label{sec:transferequation}

This section discusses the transfer equation for $\nue$ and $\nuebar$ including the \durca/ reactions in the limit of a dynamically-generated chemical potential imbalance $\beta \gg T$. The \murca/ reaction will be reviewed in Section \ref{sec:murca}. 

The electron capture reaction $e + p \leftrightarrow n + \nue$ allows the absorption and emission of $\nue$. The deviation from chemical equilibrium is measured by $\bar{\mu}_\nue \equiv \mu_e + \mu_p - \mu_n = -\beta$. 
In thermal and $\beta$ equilibrium, the neutrino distribution function is given by a Fermi-Dirac function
\be
\bar{f}_\nue & = & \left( e^{(k-\bar{\mu}_\nue)/T} + 1 \right)^{-1}.
\ee
The actual distribution function for $\nue$, denoted by $f_\nue$, is not required to have the Fermi-Dirac form here, but rather is found as a solution of the transfer equation since the weak interaction timescale is longer than the dynamical timescale in the problem.
The distribution functions for e-p-n, denoted by $f_e$, $f_p$ and $f_n$ respectively, are given by the Fermi-Dirac function since the strong and
electromagnetic interactions enforce local thermodynamic equilibrium to a good approximation.

The transfer equation for $\nue$ is given by (e.g., \citealt{1998PhRvD..58a3009R})
\be
&&  \frac{\partial f_\nue }{\partial t} + \hat{\bvec{k}} \cdot \grad f_\nue   
\nonumber \\ & = & 
2 G_{\rm F}^2 \left( c_v^2 + 3 c_a^2 \right)
\int \frac{d^3p_e}{(2\pi)^3}  \frac{d^3p_p}{(2\pi)^3} \frac{d^3p_n}{(2\pi)^3}
\nonumber \\ & \times &  (2\pi)^4 \delta^4 \left( P_p + P_e - P_n - P_\nue \right)
\nonumber \\ & \times & \left[ f_p f_e (1-f_n) (1-f_\nue) - f_n f_\nue (1-f_p) (1-f_e) \right]
\nonumber \\ & \equiv & 
j_\nue - \alpha_\nue f_\nue,
\label{eq:nuerteqn}
\ee
Here $\hat{\bvec{k}}$ is the direction of the neutrino momentum, $G_{\rm F}$, $c_v$ and $c_a$ are the usual charged-current weak interaction constants, $p_i$ is the momentum of species i, and $j_\nue$ ($\alpha_\nue$) is the emission (absorption) coefficient for $\nue$.
Small velocity dependent terms have been ignored in the matrix element.
Using energy conservation, the e-p-n Fermi-Dirac distribution functions in the absorption term can be rewritten
so that the emission and absorption coefficients are related by $\alpha_\nue \equiv j_\nue/\bar{f}_\nue$, and the right hand side of equation~(\ref{eq:nuerteqn}) becomes $\alpha_\nue(\bar{f}_\nue-f_\nue)$. In thermal and $\beta$ equilibrium, the reaction rate is then zero since $f_\nue=\bar{f}_\nue$.
The source function is defined as
\be
j_\nue & =& 2 G_{\rm F}^2 \left( c_v^2 + 3 c_a^2 \right)
\int \frac{d^3p_e}{(2\pi)^3}   
\frac{d^3p_p}{(2\pi)^3} \frac{d^3p_n}{(2\pi)^3}
\nonumber \\ & \times & (2\pi)^4 \delta^4 \left( P_p + P_e - P_n - P_\nue \right)
f_p f_e (1-f_n).
\ee
Approximations for the n-p-e integrals are discussed in \citet{1998PhRvD..58a3009R}.

The transfer equation greatly simplifies in the $\beta  \gg T $ limit. The equilibrium distribution function for $\nue$ is then
\be
\bar{f}_\nue(k) & \simeq & \Theta \left( \bar{\mu}_\nue - k \right)
= \left\{ 
\begin{tabular}{cc}
  1, & $k<\bar{\mu}_\nue$ \\
  0, & $k>\bar{\mu}_\nue$ \\
  \end{tabular}
\right.
\ee
Hence $\bar{f}_\nue(k)=0$ for $k>\bar{\mu}_\nue$, which would be true for any $k$ when $\bar{\mu}_\nue<0$.
For $\bar{\mu}_\nue > 0$ and energies $k < \bar{\mu}_\nue$, $\bar{f}_\nue(k)=1$ and the absorption coefficient $j_\nue$ equals the emission coefficient $\alpha_\nue$. 
The absorption coefficient may be computed analytically in the $T=0$ limit,
with the result
\be
\alpha_\nue & = & \frac{G_{\rm F}^2 (c_v^2 + 3c_a^2)}{4\pi^3}\, m_b^2 \mu_e \left( \bar{\mu}_\nue - k \right)^2
\Theta \left( \bar{\mu}_\nue - k \right) \Xi 
\nonumber \\ & \simeq &  
1\ \rm km^{-1}\, \left( \frac{n_e}{0.1\, n_0} \right)^{1/3} \left( \frac{\bar{\mu}_\nue}{1\, \rm MeV} \right)^2 
\nonumber \\ & \times & \left( \frac{\bar{\mu}_\nue-k}{\bar{\mu}_\nue/2} \right)^2
\Theta \left( \bar{\mu}_\nue - k \right) \Xi.
\label{eq:alphanue}
\ee
Here $m_b=938\, \rm MeV$ is the baryon mass. 
In order for either absorption or emission to occur, the step function $\Theta \left( \bar{\mu}_\nue - k \right)$ requires that $\bar{\mu}_\nue  =-\beta > 0$.
The kinematic factor $\Xi \equiv \Theta(p_{F,p}+p_{F,e} - p_{F,n})$ only allows the reaction to run at sufficiently 
large proton fraction ($x_p \ga 1/9$). 
The source function and absorption coefficient are larger at lower energies.

Similar formulae are obtained for the neutron decay reaction $n \leftrightarrow e + p + \nuebar$, which allows the absorption and emission of $\nuebar$.
The deviation from chemical equilibrium is measured by $\bar{\mu}_\nuebar \equiv \mu_n - \mu_e  - \mu_p = - \bar{\mu}_\nue = \beta $, and in the $T=0$ limit absorption and emission can only occur  when $\bar{\mu}_\nuebar = \beta > 0$.
In thermal and chemical equilibrium, and for $\beta \gg T$, the anti-neutrino distribution function is given by 
\be
\bar{f}_\nuebar & = & \left( e^{(k-\bar{\mu}_\nuebar)/T} + 1 \right)^{-1} \rightarrow \Theta \left( \bar{\mu}_\nuebar - k \right),
\ee
where $k$ is now the anti-neutrino energy. The transfer equation is
\be
&&  \frac{\partial f_\nuebar }{\partial t} + \hat{\bvec{k}} \cdot \grad f_\nuebar   
\simeq   \alpha_\nuebar (1 -  f_\nuebar ),
\label{eq:nuebarrteqn}
\ee
where $\alpha_\nuebar \simeq j_\nuebar$ and
\be
\alpha_\nuebar & \simeq & \frac{G_{\rm F}^2 (c_v^2 + 3c_a^2)}{4\pi^3}\, m_b^2 \mu_e \left( \bar{\mu}_\nuebar - k \right)^2
\Theta \left( \bar{\mu}_\nuebar - k \right) \Xi,
\label{eq:alphanuebar}
\ee
which is the same as equation~(\ref{eq:alphanue}) with $\bar{\mu}_\nue \rightarrow \bar{\mu}_\nuebar$.

Since the light crossing timescale is shorter than other timescales in the problem, the $\partial f/\partial t$ terms in equations~(\ref{eq:nuerteqn}) and (\ref{eq:nuebarrteqn}) will be ignored, with both cases leading to a static radiation transfer problem of the form
\be
\hat{\bvec{k}} \cdot \grad f &  = &  \alpha\left( 1 - f \right),
 \label{eq:staticrteqn}
\ee
where the subscript $\nue$ or $\nuebar$ has been suppressed for convenience.
In this limit the radiation field instantly adjusts to the matter distribution.
Assuming that there is no incoming intensity at the surface of the star \footnote{This ignores the emission from the other star for simplicity. Including incoming neutrinos from the other star would increase the value of $f$, which would decrease reaction rates due to additional Fermi-blocking.}, the solution to the transfer equation is
\be
f(\bvec{k},s) & = & 1 - e^{-\tau(\bvec{k},s)},
\label{eq:rtsoln}
\ee
where the optical depth from the surface over a distance $s$ is
\be
\tau(\bvec{k},s) & = & \int_0^s ds' \alpha(\bvec{k},s'),
\label{eq:tau}
\ee
and the path of integration is given by $\bvec{x}(s) = \bvec{x}(0) + s \hat{\bvec{k}}$, with the starting point for the integration $\bvec{x}(0)$ located at the surface of the star, $|\bvec{x}(0)| = R$.
There is no absorption or emission in regions where the proton fraction is too low, such that $\Xi=0$.
For each neutrino species, there is also no absorption or emission in regions where $\bar{\mu} < 0$ for that species.
Given the run of $n_e$, $n$ and $\bar{\mu}$ over the star, the $\nue$ and $\nuebar$ optical depth may
be computed. 

The scalings $j \propto \bar{\mu}^2$ in equations~(\ref{eq:alphanue}) and (\ref{eq:alphanuebar}) would seem to imply that the \durca/ rates increase strongly 
with chemical potential imbalance $\bar{\mu}$. However, for a fluid element with optical depth $\tau \gg 1$ for all neutrino directions and energies, the \durca/ reaction rates
are suppressed by a Pauli-blocking factor $1-f = e^{-\tau} \ll 1$. 
Significant reaction rates would be limited to the weakly compressed $\tau \la 1$ regions.
An estimate of the optical depth over the radius of the star is
\be
\tau & \sim  & R \alpha 
\simeq 1 
\left( \frac{R}{10\ \rm km}\right) \left( \frac{n_e}{0.1\, n_0} \right)^{1/3} \left( \frac{\bar{\mu}}{0.3\, \rm MeV} \right)^2 
\nonumber \\ & \times & \left( \frac{\bar{\mu}-k}{\bar{\mu}/2} \right)^2
\Theta \left( \bar{\mu} - k \right) \Xi,
\label{eq:taudurca}
\ee
showing that a chemical potential imbalance $\bar{\mu} \geq 0.3\, \rm MeV$ will cause the star to become optically thick to the emitted neutrinos. %Large amplitude compressions with $\bar{\mu} \ga 0.3 \, \rm MeV$ would then give large optical depths $\tau \gg 1$.

The above discussion of optical depth implies that a region undergoing large compressions $\bar{\mu} \gg 1\, \rm MeV$ and thus $\tau \gg 1$ would not be strongly heated due to the $1-f =e^{-\tau}$ suppression. Rather, it is the $\tau \sim 1$ regions is where the effects of the \durca/ reactions are largest, as they have the largest reaction rates but do not yet suffer from Fermi suppression. At large orbital separations during inspiral, 
the density perturbations and thus $\bar{\mu}$ are small. As a result, $\tau < 1$, although the \durca/ rates are nonetheless small because $\bar{\mu}$ is small. 
At small orbital separations, the density perturbations and $\bar{\mu}$ are larger. If perturbations in some region became so large that $\tau \gg 1$, the reactions 
over that region would be suppressed
by a factor $e^{-\tau}$. Note that even if the bulk of the star had $\tau \gg 1$, the regions separating compression from rarefaction will
have $\tau \sim 1$, although a small amount of mass may be contained there as compared to the bulk of the star.

\subsection{ dUrca reaction rates }
\label{sec:reactionrates}

The total reaction rates are obtained through moments of the source function over neutrino phase space.
The integrated reaction rates are needed for the evolution equation of the proton fraction.
Integrating over neutrino phase space $d^3k/(2\pi)^3$ gives the rate per unit volume
\be
n \Gamma & \equiv & \int \frac{d^3k}{(2\pi)^3}  j(1-f).
\ee
Here $\Gamma$ is the reaction rate per baryon.
In the optically thick limit where $f \rightarrow 1$, the reaction rate $n\Gamma \rightarrow 0$.
In the optically thin limit ($f\ll 1$)
\be
n \Gamma &\simeq & \int \frac{d^3k}{(2\pi)^3}  j
\nonumber \\ & =  & 
\frac{G_{\rm F}^2 (c_v^2 + 3c_a^2)}{240\pi^5}\, m^2 \mu_e  \bar{\mu}^5 \Theta(\bar{\mu}) \Xi
\nonumber \\ & = & 
5.6 \times 10^{35}\ {\rm cm^{-3}\ s^{-1}}  
\left( \frac{n_e}{n_0} \right)^{1/3} \left( \frac{\bar{\mu}}{1\, \rm MeV} \right)^5 \Xi.
\label{eq:nGamma}
\ee
This agrees with the numerical result given in equation~(18)
of \citet{1992A&A...262..131H}.

The evolution equation for the proton fraction $x_p=n_p/(n_p+n_n)$ is
\be
\frac{dx_p}{dt} & = & \Gamma_\nuebar - \Gamma_\nue,
\ee
where $\Gamma_\nuebar$ is the neutron decay rate and $ \Gamma_\nue$ is the electron capture rate.
The formulae for the two reactions are the same up to the chemical potentials $\bar{\mu}_\nuebar=-\bar{\mu}_\nue=\beta$. Only one reaction runs at any point in space, depending on whether a compression or rarefaction is occurring there. The optically thin limit simplifies to
\be
\frac{dx_p}{dt} & = & \frac{G_{\rm F}^2 (c_v^2 + 3c_a^2)}{240\pi^5n}\, m_b^2 \mu_e \Xi  \beta^5.
\ee
The  timescale to bring the system back into $\beta$-equilibrium is, for $f \ll1$, is
\be
\Gamma^{-1} & = & 400\ {\rm s}\, \left( \frac{x_p}{0.1} \right) \left( \frac{n}{n_0} \right) \left( \frac{0.1\, n_0}{n_e} \right)^{1/3} 
\left( \frac{1\, \rm MeV}{\beta} \right)^5.
\label{eq:durca_timescale}
\ee
Equation~(\ref{eq:durca_timescale}), with $n=n_0=10n_e$ and $\beta$ as shown in Fig. \ref{fig:betaT}, were used to make the equilibration timescale
shown in Fig. \ref{fig:equil_time}.
A very large imbalance ($\beta\ga 20\, \rm MeV$) is required in order for the equilibration time to becomes comparable to the orbital periods of interest ($P_{\rm orb} = 10^{-3}\, \rm s$). 
Fig. \ref{fig:betaT} shows that $\beta \la 1\, \rm MeV$ in the pre-merger phase, and hence
the equilibration time is far longer than the range of orbital period. High optical depths due to $f \simeq 1$ would 
make the equilibration timescale even longer.

As a result of the long equilibration timescale,  perturbations occur at roughly fixed proton fraction, 
with only small deviations $\Delta n_p/n_p \sim \Gamma P_{\rm orb}$ induced by the \durca/ reaction.

\subsection{ Emissivity }
\label{sec:emissivity}

The average energy of emitted neutrinos is
\be
\langle k\rangle  & \equiv & \frac{ \int d^3k  k j (1-f) }{\int d^3k  j (1-f)}.
\ee
In the optically thin limit, $f \ll 1$, this approaches
\be
\langle k\rangle  & \simeq &
 \frac{\int_0^{\bar{\mu}} dk k^3 \left( \bar{\mu}-k \right)^2}{
\int_0^{\bar{\mu}} dk k^2 \left( \bar{\mu}-k \right)^2 } = \frac{1}{2}   \bar{\mu}.
\label{eq:kavg}
\ee
In the optically thick limit, since the source function $j \propto (\bar{\mu}-k)^2$, it is expected that small $k < \bar{\mu}$ become optically thick first.
This will weight the integrals toward larger $k$, and hence $\bar{\mu}/2 \la \langle k\rangle \la \bar{\mu}$ in general.
This fact will be important in Section \ref{sec:entropy} in the entropy evolution equation.

The emissivity, i.e., energy emission rate, per unit volume and per unit time, is
\be
\varepsilon & \equiv & \int \frac{d^3k}{(2\pi)^3}  k j (1-f)   \equiv  n \Gamma \langle k \rangle.
\ee
In the optically thick limit, $f \rightarrow 1$, the emission is suppressed. In the optically thin limit, $f \ll 1$, 
\be
\varepsilon & \simeq & \frac{1}{2} n\Gamma \bar{\mu} = 
\frac{G_{\rm F}^2 (c_v^2 + 3c_a^2)}{480\pi^5}\, m^2 \mu_e  \bar{\mu}^6 \Theta(\bar{\mu}) \Xi
\nonumber \\ & = & 4.5 \times 10^{29}\, {\rm erg\ cm^{-3}\ s^{-1}}\, 
\left( \frac{n_e}{n_0} \right)^{1/3} \left( \frac{\bar{\mu}_i}{1\, \rm MeV} \right)^6 \Xi.
\label{eq:varepsilon}
\ee
This agrees with the numerical result given in equation~(18)
of \citet{1992A&A...262..131H} which has been used in equation \ref{eq:varepsilon_H92}.

\subsection{ Net heating and total energy loss}
\label{sec:entropy}

Neutrino emission cools the star while the release of Gibbs energy by reactions heats the star. The difference of these two quantities gives the net heating
rate, while the sum is the total damping rate of mechanical energy.

The npe$\mu$ gas entropy equation \citep{1986bhwd.book.....S} including these opposing effects is
\be
n T \frac{ds}{dt} & =& - \grad \cdot \bvec{F}_\nue - \sum_{i=n,p,e} \mu_i \frac{dn_i}{dt} \rfloor_\nue
\nonumber \\ & - & 
  \grad \cdot \bvec{F}_\nuebar - \sum_{i=n,p,e} \mu_i \frac{dn_i}{dt} \rfloor_\nuebar
\nonumber \\ & =& 
- \grad \cdot \bvec{F}_\nue + n\Gamma_\nue  \bar{\mu}_\nue
-  \grad \cdot \bvec{F}_\nuebar+ n\Gamma_\nuebar  \bar{\mu}_\nuebar,
\label{eq:sdot1}
\ee
Here $s$ is the entropy per baryon.
Two kinds of processes change the gas entropy. The first type of interaction is with an ``external system", the neutrinos, which exchanges energy with the gas through a divergence of the neutrino energy flux, $\grad \cdot \bvec{F}$. The second kind of change of entropy is internal to the system composed of the npe$\mu$ gas, and is given by the change in Gibbs energy of npe$\mu$ due to the Urca reaction. 

In the static limit, the divergence of the energy flux is found by integrating equation~(\ref{eq:staticrteqn}) against $k d^3k/(2\pi)^3$, giving
\be
\grad \cdot \bvec{F} & = &  \varepsilon
\label{eq:divF}
\ee
where the energy flux is defined by 
\be
\bvec{F} & =& \int \frac{d^3k}{(2\pi)^3} \bvec{k} f.
\ee
Plugging this result into equation~(\ref{eq:sdot1}) gives (e.g. \citealt{1988PhR...163...95C})
\be
n T \frac{ds}{dt}  & = & 
n \Gamma_\nue \left( \bar{\mu}_\nue - \langle k_\nue \rangle \right)
+ n \Gamma_\nuebar \left( \bar{\mu}_\nuebar - \langle k_\nuebar \rangle \right).
\label{eq:entropy_thick}
\ee
The result in equation~(\ref{eq:entropy_thick}) is valid for arbitrary optical depths. Since $\bar{\mu}/2 \leq \langle k \rangle \leq \bar{\mu}$, this equation shows that the entropy {\it always increases} in the $\beta \gg T$ limit. That is, in the $\beta \gg T$ limit, the chemical heating term is always larger than the neutrino cooling term. 
In the optically thin limit, $\langle k \rangle = \bar{\mu}/2$, and the chemical heating term is twice the cooling term and the net heating rate is
\be
n T \frac{ds}{dt}  & = & 
\frac{1}{2} n \Gamma_\nue \bar{\mu}_\nue 
+ \frac{1}{2} n \Gamma_\nuebar \bar{\mu}_\nuebar
\nonumber \\ & = & 
\frac{G_{\rm F}^2 (c_v^2 + 3c_a^2)}{480\pi^5}\, m_b^2 \mu_e \Xi  \beta^6,
\label{eq:entropy_thin}
\ee
which is equal to the emissivity in equation~(\ref{eq:varepsilon}).
%Qualitatively, \nevin{why only qualitatively?} this agrees with the result of FW, who showed that in the nonlinear limit $\beta \gg T$, the heating associated with Urca reactions dominates the neutrino cooling. 
The factor $1/2$ in equation~(\ref{eq:entropy_thin}) is for the \durca/ reactions, while FW found a factor $5/8$ for \murca/ (Section \ref{sec:murca}).

\subsection{ mUrca reactions }
\label{sec:murca}

For the \murca/ reactions,  the rate per volume is \citep{1992A&A...262..131H} 
\be
n\Gamma & = & 1.6 \times 10^{29}\, {\rm cm^{-3}\ s^{-1}}\, \left( \frac{n_e}{n_0} \right)^{1/3}\left( \frac{\bar{\mu}}{1\, \rm MeV} \right)^7
\ee
and the emissivity is
\be
\varepsilon & =& 9.6 \times 10^{22}\, {\rm erg\, cm^{-3}\, s^{-1}}\, \left( \frac{n_e}{n_0} \right)^{1/3}\left( \frac{\bar{\mu}}{1\, \rm MeV} \right)^8.
\ee
The integral for the rate contains the dependence $n\Gamma \propto \int dk k^2 (k-\bar{\mu})^4$. 
Using these results, the source function is found to be
\be
j & \simeq& \alpha \simeq  10^{-3}\, \rm km^{-1}\, \left( \frac{n_e}{n_0} \right)^{1/3} \left( \frac{\bar{\mu}-k}{1\, \rm MeV} \right)^4 \Theta(\bar{\mu}-k).
\label{eq:jMURCA}
\ee
Note that there is no $\Xi$ factor as the bystander neutron allows this reaction to run with no restriction to large proton fraction. 
The mean neutrino energy in the $f \ll 1$ limit is then $\langle k \rangle = 3\bar{\mu}/8$.
The absorption coefficient, rate and emissivity are orders of magnitude smaller than for \durca/ for $\beta \sim \rm MeV$. 

The \murca/ $n-p-e$ entropy equation, including both neutrino cooling and
chemical heating, gives
\be
n T \frac{ds}{dt} & = & - n \left( \frac{3}{8} \bar{\mu}_\nue \Gamma_\nue + \frac{3}{8} \bar{\mu}_\nuebar \Gamma_\nuebar \right) 
+ n \left( \bar{\mu}_\nue \Gamma_\nue + \bar{\mu}_\nuebar \Gamma_\nuebar \right)
\nonumber \\ & = & 
\frac{5}{8} n \left( \bar{\mu}_\nue \Gamma_\nue + \bar{\mu}_\nuebar \Gamma_\nuebar \right)
\label{eq:murca_opt_thin}
\ee
in the optically thin regime. The $3/8$ factor for the neutrino cooling and the $5/8$ factor for the net heating agree with FW. 

The derivation of the \murca/ net heating rates given by FW differs from that given here. Both treatments assume an adiabatic density perturbation $\Delta \rho/\rho$ and ignore the effect of reactions on the density perturbation. FW then compute the non-adiabatic pressure perturbation, $\Delta p$, due to the reactions. This pressure perturbation is averaged over a pulsation cycle to compute the  $\int \Delta p dV$ work, and this was treated as the chemical heating rate of the gas. FW then subtract the neutrino cooling from that expression to find the net heating, which is in agreement with our equation~(\ref{eq:murca_opt_thin}). The present derivation, using equation~(\ref{eq:sdot1}), gives the same result as FW in the optically thin limit. Further, the effect of finite optical depth is also included in the present formulation.

\subsection{ Results }
\label{sec:thickresults}

\begin{figure}
%\epsscale{2.5}
%\plotone{Edot_vs_nu_1p4_1p4_thick.pdf}
\includegraphics[width=\columnwidth]{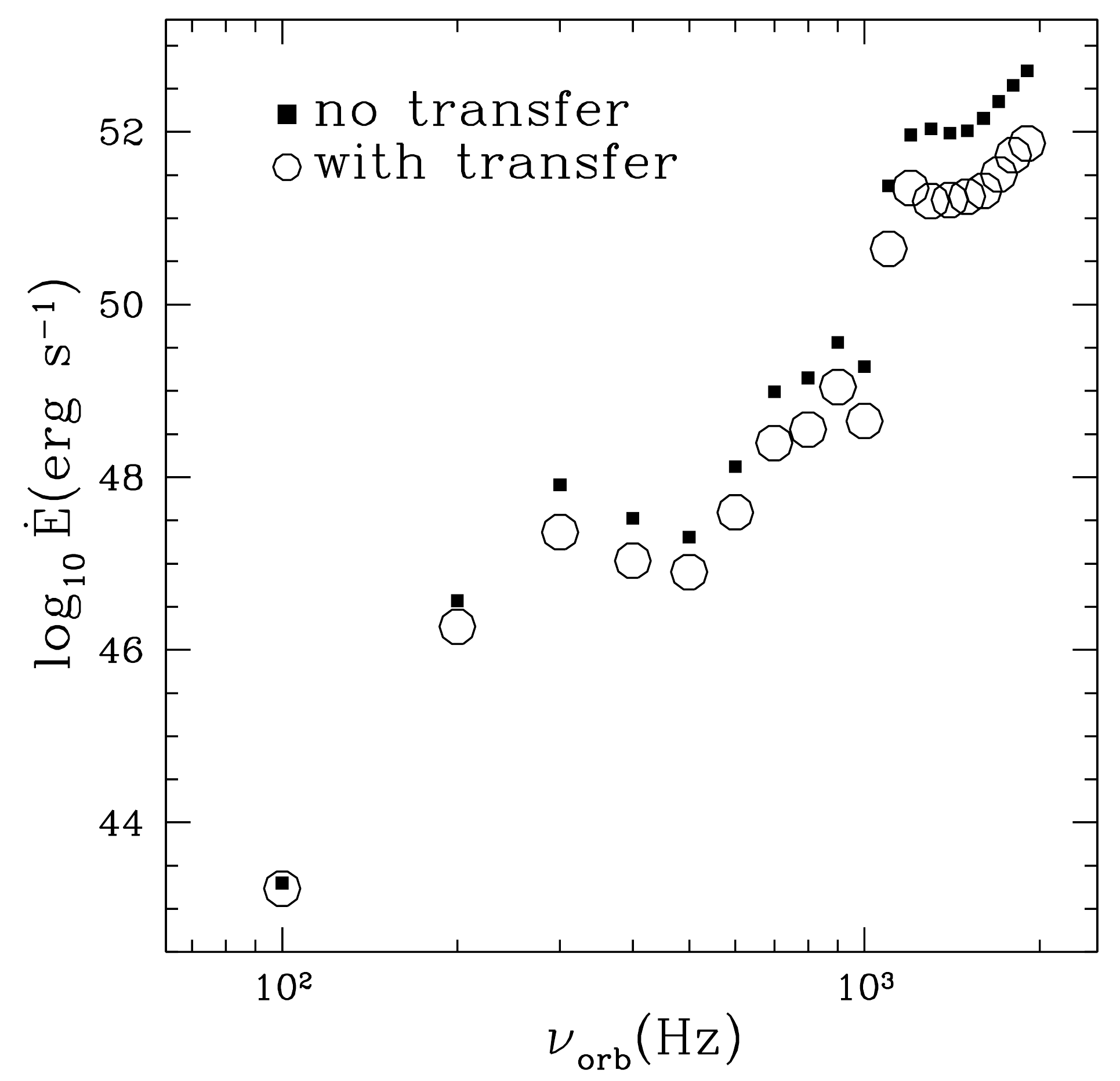}
\caption{dUrca heating rate versus orbital frequency, in which we compare the optically thin approximation (solid squares) to the calculation including radiation transfer effects (open circles). Results are presented for points spaced every $100\, \rm Hz$ in orbital frequency.   }
\label{fig:Edotthick}
\end{figure}

Fig. \ref{fig:Edotthick} compares the optically thick heating rate to the optically thin rate already shown in Fig. \ref{fig:Edot}. Since the optically thick calculations are more time consuming, heating rates are only computed every $100\, \rm Hz$ in orbital frequency.
The optically thick rate was computed by integrating equation~(\ref{eq:entropy_thick}) over the star. This was carried out in spherical coordinates using the density perturbations $\Delta \rho(r,\theta,\phi,t)$ constructed from the mode amplitudes and eigenfunctions, as well as the optical depths $\tau(r,\theta,\phi,k,\hat{\gv{k}},t)$. By contrast, the optically thin calculation in equation~(\ref{eq:Edotheatlm}) uses a tabulated integral over the six spherical harmonics with a numerical integral over $r$. 

At wide orbital separation (small orbital frequency $\la 100\, \rm Hz$), the optically thick and thin calculations in Fig. \ref{fig:Edotthick} agree. Once the g$_1$ mode has been excited at $\nu_{\rm orb} \ga 200\, \rm Hz$, the chemical potential imbalance and optical depth are large enough for Fermi-blocking effects to become noticeable. Keeping in mind the large oscillations seen in Fig. \ref{fig:Edot}, there is a general trend that the optically thick heating rate is a factor of $\sim 2-10$ smaller than the optically thin heating rate over the range $200 < \nu_{\rm orb} < 1000\, \rm Hz$. Interestingly, the rate continues to increase at (unphysically large) $\nu_{\rm orb}=1000-2000\rm Hz$, despite reaching maximum values of $\beta \sim 10\, \rm MeV$ and thus large optical depths. This is because there are still spatial regions and neutrino directions for which $f \la 1$ and $\tau \la 1$.
 
The optically thick calculation computes the neutrino distribution function at a grid of spatial points in the star, and at a grid of photon directions $\hat{\gv{k}}$. The distribution function at the surface of the star can be used to compute the total specific neutrino luminosity 
\be
L_E & = & R^2 \int d\Omega \int \frac{k^3 d\Omega_k}{(2\pi)^3} f(R,\theta,\phi,k,\hat{\bvec{k}}) \Theta( \gv{n} \cdot \hat{\gv{k}} )
\label{eq:L_E}
\ee
where $\gv{n}$ is the radial direction, $d\Omega$ is the solid angle integral over the surface of the star, $d\Omega_k$ is the solid angle ntegral over the neutrino directions, and the $\Theta( \gv{n} \cdot \hat{\gv{k}} )$ factor only allows outgoing directions in the integral.

\begin{figure}
%\epsscale{2.5}
%\plotone{spectrum_1p4_1p4.pdf}
\includegraphics[width=\columnwidth]{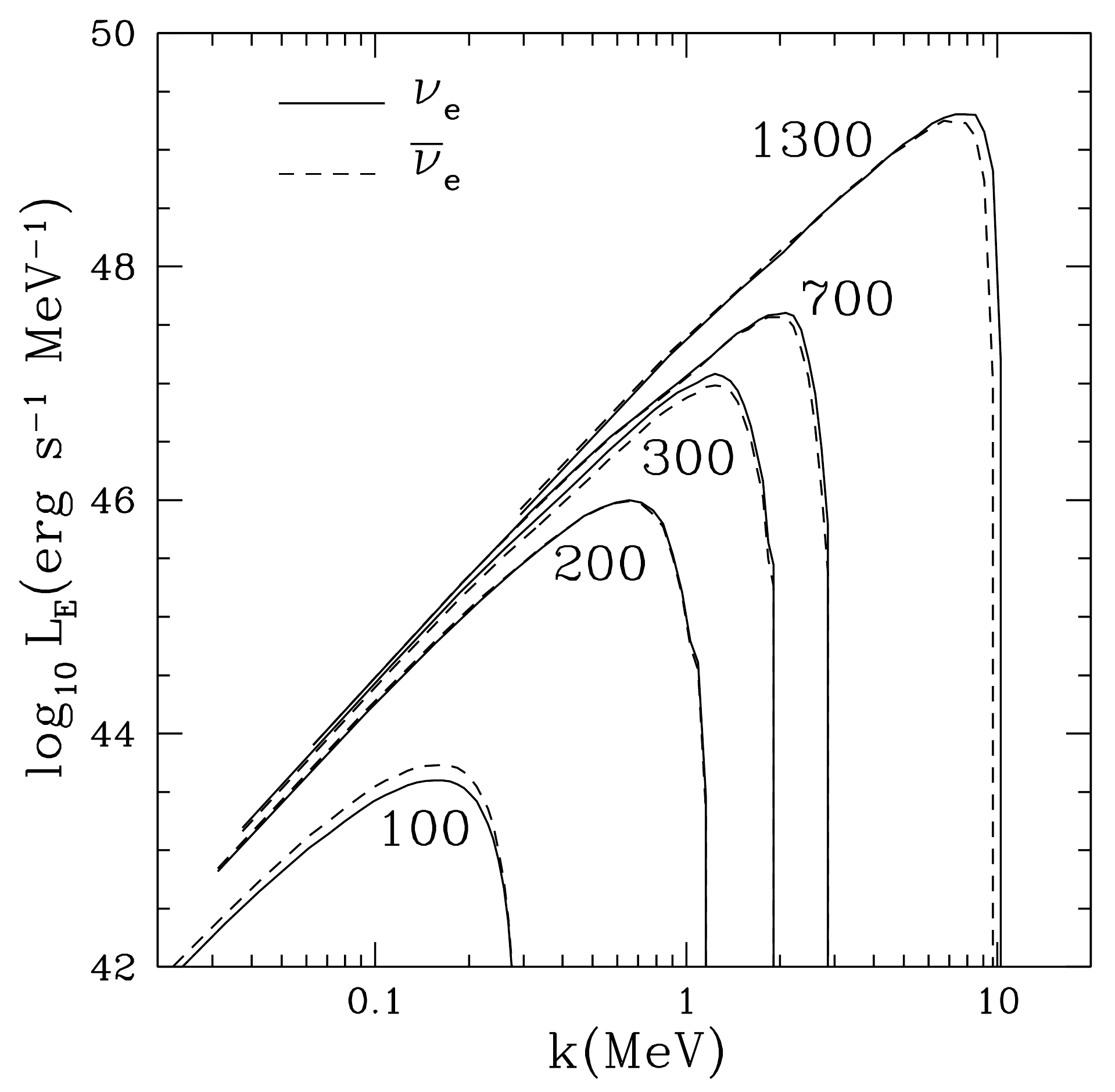}
\caption{dUrca specific luminosity ($L_E$) versus neutrino energy ($k$) at different points during the inspiral, as labeled by orbital frequency in Hz. The solid line represents $\nu_e$ while the dashed line represents $\bar{\nu}_e$.}
\label{fig:spectrum}
\end{figure}

The specific luminosity in equation~(\ref{eq:L_E}) has been computed both for $\nue$ and $\nuebar$. The resulting spectra are shown in Fig. \ref{fig:spectrum} for a range of orbital periods. Since $\beta$ is larger at smaller orbital separation, the emission rate increases and the neutrino spectrum hardens. Near merger, we find $L_E \simeq 10^{49}\, \rm erg\ s^{-1}\, MeV^{-1}$ for neutrinos with energy $k \sim \beta \sim \rm a\, few\, MeV$.

\section{ discussion }
\label{sec:discussion}

This paper considered Urca reactions in the neutron star core driven by tidally induced compression during binary neutron star inspiral. Urca reactions operate when fluid compression drives the matter out of $\beta$-equilibrium, and the rates are insensitive to temperature when $\kB T \ll \beta$. Motivation for this work came from equations~(\ref{eq:Edotheat}) and (\ref{eq:varepsilon_H92}), which show that a chemical potential imbalance $\beta \sim 10\, \rm MeV$ will lead to a dUrca heating rate and neutrino luminosity $\sim 10^{53}\, {\rm erg\ s^{-1}}$, comparable to that found from the post-merger accretion disk or core-collapse supernovae (e.g. \citealt{2011PhRvL.107e1102S}). 
If present, such large neutrino emission can be important for driving an outflow through heating of the gas or through $\nu-\bar{\nu}$ annihilation
\citep{2009ApJ...690.1681D} and also through changing the electron fraction $Y_e$ of unbound gas, affecting nucleosynthesis (e.g. \citealt{2015ApJ...815...82L}).
For thermodynamic derivative $\betan = (\partial \beta/\partial \ln \rho)_{s,x} \sim 100\, \rm MeV$ (see figures~\ref{fig:model} and \ref{fig:beta0_EOS}), $\beta \simeq \betan \Delta \rho/\rho \sim 10\, {\rm MeV}$ requires a compression of only $\Delta \rho/\rho \sim 0.1$. However our model does not find such large compressions during the inspiral, and due to the $\beta^6$ dependence of the dUrca rates, the pre-merger emission is found to be much smaller than the post-merger emission from the accretion disk.

Our calculations use direct and modified Urca reactions operating in a $M=1.4\, M_\odot$ neutron star,  which is constructed by solving the TOV equation using the Skyrme Rs equation of state. At close separation, the  compression arising from the excitation of the quadrupolar g$_1$ and f-modes drive chemical potential imbalances of $\beta \sim \rm 1-5\, MeV$ (see Figure~\ref{fig:betaT}). If the stellar mass and equation of state allow sufficiently high proton fraction that dUrca reactions operate, and if the conditions are optically thin, then heating prior to merger may produce a temperature $\kB T \sim 10^{-2}\, \rm MeV$. This core-averaged temperature is larger than that produced by shear viscosity \citep{1994MNRAS.270..611L} by an order of magnitude, but much smaller than some calculations which assumed efficient tidal dissipation (e.g. \citealt{1992ApJ...397..570M}). Hence this work reinforces the idea that the neutron stars are cold at merger and that tides are inefficient at transferring energy and angular momentum from the orbit to the star.

Near merger, even at $T=0$, highly compressed and rarefied regions become optically thick and neutrino Fermi-blocking affects the Urca reaction rates. An approximate solution to the full three-dimensional transfer problem was given to estimate the size of this effect. It was found that neutrino Fermi-blocking during the inspiral decreases the heating rates by a factor of a few, which would give slightly lower core-averaged temperatures.

The static transfer equation~(\ref{eq:staticrteqn}) and its solution as given by equations~(\ref{eq:rtsoln}) and ~(\ref{eq:tau}) are solved in Section \ref{sec:thickresults} over a grid of neutrino direction and energy, as well as over a spatial grid for the background star. This solution differs from commonly used leakage schemes for hot matter (e.g. \citealt{2003MNRAS.342..673R}) which interpolate between the optically thin and diffusion approximations. Our solution does not assume the neutrino distribution function is nearly Fermi-Dirac, but rather it is found as a solution to the transfer equation.

Assuming n-p-e-$\mu$ matter, the largest uncertainty in our results is due to the equation of state of nuclear matter at high densities. If proton fractions are low and dUrca reactions are not possible, mUrca reactions give neutrino emission and heating rates orders of magnitude smaller for $\beta \sim 1-10\, \rm MeV$. If high proton fractions are only possible for large neutron star masses, this may require one or both of the stars in the binary to have sufficiently large mass to allow dUrca reactions. Another uncertainly in our results is the simplification of using a TOV solution for the background model but Newtonian equations for the fluid perturbations and neutrino transfer. This implies the background model is effectively not in Newtonian hydrostatic balance. A consistent use of General Relativistic solutions for the background and perturbations could find significant deviations from the results here.  As discussed by \citet{1994ApJ...432..296R}, the mode overlap integrals could be different by a factor of $\sim 2$. This would imply dUrca heating rates different by $\sim 30$. Lastly,  General Relativistic solutions of the neutrino transfer equation would include changes in neutrino energy due to redshift and fluid motions.

In this paper, heating by the Urca reactions is discussed in the language of ``chemical energy" added through reactions as well as energy lost from the n-p-e-$\mu$ gas to neutrino radiation. The Urca reactions only affect the forces in the momentum equation indirectly, through reactions changing the density and pressure as compared to what they would be for adiabatic motions with no reactions. We agree with \citet{1993A&A...271..187G} that this description is simpler and more natural than that of bulk viscosity damping of compressive motions. The use of a bulk viscosity implies a viscous force in the momentum equations as well as a viscous heating term in the energy equation. In the limit $\beta \ll T$, \citet{1989PhRvD..39.3804S} derived a frequency-dependent bulk viscosity for fluid undergoing harmonic compressive motions. Such a description cannot be used in a time-dependent hydrodynamics code with the fluid not undergoing strictly harmonic motion. A second issue is that the bulk viscous heating must always be combined with neutrino cooling to determine the temperature evolution. In the $\beta \ll T$ limit, net cooling always occurs. Only in the nonlinear limit of $\beta \gg T$ does net heating occur (FW), but the ``linear" bulk viscosity derived by \citet{1989PhRvD..39.3804S} is not valid in that limit, and underestimates the heating. It is in principle possible to describe the $\beta \gg T$ limit as a non-Newtonian fluid with a bulk viscosity nonlinear in the compression. Lacking any knowledge of microphysical processes, a non-Newtonian parametrization of bulk viscosity is useful as a phenomenological model. But the Urca reactions allow a microphysical understanding and so the language of bulk viscosity is not required. Lastly, if one is already solving the equations of hydrodynamics, composition changes due to Urca reactions, and some approximation of neutrino radiation transfer, there is no need to insert an additional bulk viscous force into the momentum equation and bulk viscous heating into the energy equation (e.g. \citealt{2001A&A...380..544R}) as the Urca reactions are already being taken into account through the composition and energy equations.

\citet{1993A&A...271..187G} found $\beta \sim k \sim T \sim  10\, \rm MeV$ generated during collapse of a cold neutron star to a black hole. In that work, the emergent antineutrinos suffer from gravitational redshift as well as Doppler redshifting due to the infall. They mention that antineutrino absorption may become important. The dUrca optical depth in our equation~(\ref{eq:taudurca}) is $\tau \sim 1\, (n_e R^3/1.6\times10^{55})^{1/3} (\beta/{\rm 0.3\, MeV})^2$. During collapse, $n_e R^3 \sim \rm constant$, but $\beta$ increases significantly. The optical depth is therefore expected to increase, leading to $f_\nuebar \sim 1$ and Fermi-blocking in the optically thick regions. Since this will limit the rates of heating and neutrino emission in optically thick regions, we expect $T \ll 10\, \rm MeV$ and neutrino emission to infinity limited by $f_\nuebar \la 1$.

Large amplitude oscillations in the core of the merger remnant may produce a detectable gravitational wave signal. For long-lived merger remnants, the damping time of the oscillations may be set by a combination of gravitational wave emission, hydrodynamic shocks and nonlinear wave interaction, or by Urca processes. Core temperatures $T \gg 1\, \rm MeV$ have been found due to shock heating during the merger \citep{2011PhRvL.107e1102S}. Hence depending on the size of the chemical potential imbalance $\beta$ created by the oscillation, the $\beta \gg T$ limit used here may not be appropriate for the Urca rates, and the finite temperature rates in \citet{1992A&A...262..131H} must be used. However, the high temperatures and large $\beta$ imply optically thick conditions to both charged and neutral current reactions in the remnant core, and optical depth effects as described here must be included, likely giving a large suppression of damping rates compared to the optically thin case. For the example given in Fig. \ref{fig:Edotthick}, the $|q_f|\sim1$ f-mode oscilllation at $\nu_{\rm orb}=2000\, \rm Hz$  has energy $\sim 4\times 10^{53}\, \rm erg$ and dUrca damping rate $\sim 10^{52}\rm erg\ s^{-1}$, leading to a damping time $\sim 40\, \rm s$, and hence may be comparable to or longer than the cooling time of the remnant. This damping time is in rough agreement with that found by \citet{1991ApJ...373..213I} if a temperature $T \sim 10\, \rm MeV$ is used in their formula. For the f-mode, the gravitational wave damping time is  $\simeq 10^{-2}-10^{-1}\rm s$ \citep{1991ApJ...373..213I}, and hence dUrca damping is likely unimportant. If the merger remnant contains modes with smaller quadrupole moments than the f-mode (e.g. g-modes) which have longer gravitational wave damping times \citep{1994ApJ...426..688R}, the amplitude-dependent damping rates discussed here may be relevant in setting the mode damping timescales.

%\nevin{is there anything we can mention about post-merger $\beta$-disequilibrium? Post-merger oscillations out of beta equilibrium. f-mode ringing in post-merger star. damping time? if order unity or 10\% density perturbation after merger. \citet{1993A&A...271..187G} discuss capture of neutrinos by coherent scattering in the crust. estimate mfp, and enough neutrinos to change $Y_e$ before merger. temp is 0.01MeV, but neutrino emission is based on beta. they show mfp of large nuclei is short. }

%\phil{ Numerical simulations??? } Review paper by Vaiotti and Rezzola on binary neutron star mergers. They have a section on neutrino transport. Shibata group uses \citet{2010PThPh.124..331S}

%\acknowledgements
\section*{Acknowledgements}
We are grateful to Shane Davis, Kent Yagi and Hang Yu for useful conversations. This research was supported by NASA grant NNX14AB40G.

%\appendix

\bibliographystyle{mnras}
\bibliography{ref}

\bsp	% typesetting comment
\label{lastpage}
\end{document}